\documentclass[a4paper,11pt]{article}
\usepackage[english]{babel}
\usepackage{jheppub}
\pdfoutput=1
\usepackage[T1]{fontenc}
\usepackage{bbold}
\usepackage{float}
\usepackage{amsmath}
\usepackage{empheq}
\usepackage{booktabs}
\usepackage{color}
\usepackage[utf8]{inputenc}
\usepackage{xspace}
\usepackage{scalerel}
\usepackage[most]{tcolorbox}
\usepackage{multirow}
\usepackage{scalefnt}
\usepackage{bold-extra}
\usepackage[shortlabels]{enumitem}
\usepackage[tikz]{bclogo}
\usepackage{subcaption}
\usepackage{tikz-feynman}
\usepackage{cancel}
\usepackage{verbatim}
\usetikzlibrary{arrows,shapes}
\usepackage{afterpage}
\usepackage{graphicx,grffile}
\usepackage{soul}

\newcommand{\stepone}{{Step\,I}}
\newcommand{\steptwo}{{Step\,II}}
\newcommand{\stepthree}{{Step\,III}}

\newcommand\F{${\rm F}$}
\newcommand\FJ{${\rm FJ}$}
\newcommand\FJJ{${\rm FJJ}$}

\newcommand\ww{$W^+W^-$}
\newcommand\zz{$ZZ$}

\newcommand{\as}{\alpha_s}

\newcommand{\pt}{{p_{\text{\scalefont{0.77}T}}}}

\newcommand{\ptrad}{{p_{\text{\scalefont{0.77}T,rad}}}}

\newcommand{\pte}{{p_{\text{\scalefont{0.77}T,$e^-$}}}}

\newcommand{\ptlone}{{p_{\text{\scalefont{0.77}T,$\ell_1$}}}}

\newcommand{\muF}{{\mu_{\text{\scalefont{0.77}F}}}}
\newcommand{\muR}{{\mu_{\text{\scalefont{0.77}R}}}}

\newcommand{\Q}{{Q_{\text{\scalefont{0.77}$0$}}}}

\newcommand{\Qres}{{Q_{\text{\scalefont{0.77}res}}}}

\newcommand{\noun}[1]{{\scshape #1}}

\newcommand{\POWHEG}{\noun{Powheg}}

\newcommand{\POWHEGBOX}{\noun{Powheg-Box}}
\newcommand{\POWHEGBOXRES}{\noun{Powheg-Box-Res}}

\newcommand{\minlo}{{\noun{MiNLO$^{\prime}$}}\xspace}
\newcommand{\minnlo}{{\noun{MiNNLO$_{\rm PS}$}}\xspace}

\newcommand{\Matrix}{{\noun{Matrix}}\xspace}
\newcommand{\OpenLoops}{{\noun{OpenLoops}}\xspace}
\newcommand{\Recola}{{\noun{Recola}}\xspace}
\newcommand{\PYTHIA}[1]{\noun{Pythia{#1}}\xspace}
\newcommand{\radish}{\textsc{RadISH}\xspace}

\newcommand{\setupinclusive}{{\tt setup-inclusive}\xspace}
\newcommand{\setupfiducial}{{\tt setup-fiducial}\xspace}

\newcommand{\citere}[1]{ref.\,\cite{#1}}

\newcommand{\citeres}[1]{refs.\,\cite{#1}}

\newcommand{\eqn}[1]{eq.\,(\ref{#1})}

\newcommand{\fig}[1]{figure\,\ref{#1}}

\newcommand{\tab}[1]{table\,\ref{#1}}
\newcommand{\sct}[1]{section~\ref{#1}}

\newcommand{\LambdaPWG}{\Lambda_{\rm pwg}}

\newcommand{\sphid}[1]{}

\usepackage{xcolor}

\newcommand{\tmop}[1]{\ensuremath{\operatorname{#1}}}

\newtcolorbox{empheqboxed}{colback=white!35, 
 colframe=black,
 width=\textwidth,
 sharpish corners,
 top=-2mm, 
 bottom=0pt
}

\title{{\boldmath{$ZZ$} production at nNNLO+PS with M{\scalefont{0.77}I}NNLO\boldmath{$_{\text{PS}}$}}}

\author[]{Luca Buonocore,$^{a}$}
\author[]{Gabri\"el Koole,$^{b}$}
\author[]{Daniele Lombardi,$^{b}$}
\author[]{Luca Rottoli,$^{a}$}
\author[]{\quad\quad \quad\quad \quad\quad Marius Wiesemann$^{b}$}
\author[]{and Giulia Zanderighi$^{b}$}

\emailAdd{buonocore@uzh.ch}
\emailAdd{koole@mpp.mpg.de}
\emailAdd{lombardi@mpp.mpg.de}
\emailAdd{rottoli@uzh.ch}
\emailAdd{wieseman@mpp.mpg.de}
\emailAdd{zanderi@mpp.mpg.de}

\affiliation[]{$^{a}$University of Zurich, Winterthurerstrasse 190,
  8057 Zurich, Switzerland}

\affiliation[]{$^{b}$Max-Planck-Institut f\"ur Physik, F\"ohringer
  Ring 6, 80805 M\"unchen, Germany}

\abstract{We consider $ZZ$ production in hadronic collisions and
  present state-of-the-art predictions in QCD perturbation theory
  matched to parton showers. Next-to-next-to-leading order corrections
  to the quark-initiated channel are combined with parton
  showers using the \minnlo{} method, while next-to-leading order
  corrections to the loop-induced gluon fusion channel are matched using the
  \POWHEG{} method.  Their combination, dubbed nNNLO+PS, constitutes
  the best theoretical description of $ZZ$ events to date. Spin
  correlations, interferences and off-shell effects are included by
  calculating the full process $pp \to
  \ell^+\ell^-\ell^{(\prime)+}\ell^{(\prime)-}$.  We show the crucial
  impact of higher-order corrections for both quark- and
  gluon-initiated processes as well as the relevance of the parton
  shower in certain kinematical regimes. Our predictions are in very
  good agreement with recent LHC data.}


\preprint{
\vspace{-24pt}
  \begin{flushright}
    MPP-2021-130\\
    ZU-TH 36/21
  \end{flushright}
}

\begin{document}

\maketitle

\section{Introduction}
\label{sec:intro}

Vector-boson pair production processes provide some of the most
relevant signatures in precision measurements, which have evolved to one
of the cornerstones of the rich physics programme at the Large Hadron
Collider (LHC). The accurate determination of production rates and
distributions provides a valuable path towards the observation of
deviations from the predictions made by the Standard Model (SM) of
particle physics.  Observing or constraining anomalous interactions
among SM particles is one of the central goals of such
analyses. Through diboson signatures the couplings among three vector
bosons (triple-gauge couplings) are directly accessible, which are
altered by various beyond-the-SM (BSM) theories. Therefore, the
observation of small deviations from the expected rates or shapes of
distributions would be a clear sign of new physics.  Similarly,
measurements at high transverse momentum of some of the particles
produced in diboson processes provide constraints on the mass range of
possible heavy $Z'$ bosons.  Apart from that, vector-boson pair final
states constitute an irreducible background to on- and off-shell Higgs
cross-section measurements, when the Higgs boson decays to four
leptons. These measurements are important for the extraction of the
Higgs couplings and for constraints of the Higgs width
\cite{Caola:2013yja, Campbell:2013una, Campbell:2013wga, Ellis:2014yca,Grazzini:2021iae,Khachatryan:2014iha,Aad:2015xua,Khachatryan:2015mma,Khachatryan:2016ctc,Aaboud:2018puo,Sirunyan:2019twz}.

While the cross section for \zz{} production is smaller than the one
of $W^\pm Z$ or $W^+W^-$ production, experimentally the decay to four
charged leptons provides the cleanest signature of the massive diboson
processes, since the final state does not involve any missing
transverse momentum.  Accordingly, experimental measurements already
reach a remarkable level of precision.  In particular, both ATLAS and
CMS collaborations have performed measurements of the $ZZ$ production
cross sections at 5.02 TeV~\cite{CMS:2021pqj}, 7
TeV~\cite{Aad:2011xj,Aad:2012awa,CMS:2012exm,CMS:2015qgb,Khachatryan:2015pba},
8
TeV~\cite{CMS:2013piy,CMS:2014xja,CMS:2015qgb,ATLAS:2015rsx,Khachatryan:2015pba,Aaboud:2016urj,CMS:2018ccg}
and 13
TeV~\cite{CMS:2016ogx,Aaboud:2017rwm,CMS:2017dzg,CMS:2018ccg,Aaboud:2019lgy,CMS:2020gtj}
and used these measurements to test the SM and constrain triple-gauge
couplings.

Modern fits of parton distribution functions (PDFs) have started to
include more and more LHC data. For instance
NNPDF3.1~\cite{Ball:2017nwa} already includes top-pair production and
the transverse momentum of the Drell-Yan pair. Upcoming fits will also
include direct photon, dijet and single top production.  It is clear
that a further step would be the inclusion of diboson production and
other processes with more final-state particles such as three jets,
provided the accuracy of theory predictions, both at the level of
higher-order QCD and electroweak (EW) corrections, is sufficient.

The first next-to-leading order (NLO) QCD corrections to $Z$-boson
pair production started to appear about thirty years
ago~\cite{Mele:1990bq,Ohnemus:1990za,Ohnemus:1994ff,Dixon:1998py,Campbell:1999ah}. NLO
QCD calculations were later consistently matched with fully exclusive
Parton Shower Monte Carlo programmes (NLO+PS) using the
\POWHEG{}~\cite{Melia:2011tj,Nason:2013ydw} or aMC@NLO
method~\cite{Frederix:2011ss}.  Electroweak (EW) effects at NLO were
also computed first in the on-shell
approximation~\cite{Bierweiler:2013dja,Baglio:2013toa} and
later keeping off-shell and spin-correlation
effects~\cite{Biedermann:2016yvs,Biedermann:2016lvg}.
The combination of NLO QCD and NLO EW corrections was presented in
\citere{Chiesa:2018lcs} and recently also their matching to parton
showers was performed \cite{Chiesa:2020ttl}. Likewise, in the case of
polarized $Z$ bosons NLO QCD and NLO EW corrections have been combined
very recently \cite{Denner:2021csi}.
NNLO QCD corrections have been computed for on-shell
\cite{Cascioli:2014yka,Heinrich:2017bvg} and off-shell \zz{}
production~\cite{Grazzini:2015hta,Kallweit:2018nyv}, and their
combination with NLO EW effects was presented
in~\citere{Kallweit:2019zez}.
The loop-induced $gg \to ZZ$ process starts contributing only at ${\cal
  O}(\as^2)$, but it is enhanced by the gluon PDFs. Since 
higher-order corrections to this
process can be formulated as a gauge-invariant set of contributions and
their impact was expected to be important,
NLO QCD corrections to $gg \to ZZ$ production have also been computed
in the recent years~\cite{Caola:2015psa,Caola:2016trd,Grazzini:2018owa,Grazzini:2021iae}. 
The leading order (LO) matching of the loop-induced gluon fusion process was 
presented in \citere{Binoth:2008pr}, while NLO+PS predictions were first obtained neglecting
the quark channels \cite{Alioli:2016xab} and very recently also including the full NLO QCD
corrections with quark-gluon and quark-antiquark channels and the Higgs resonance \cite{Alioli:2021wpn}. 

The remarkable progress in NNLO QCD calculations\footnote{By now all $2 \to 1$ and $2 \to 2$ colour-singlet processes are available at NNLO QCD \cite{Ferrera:2011bk,Ferrera:2014lca,Ferrera:2017zex,Campbell:2016jau,Harlander:2003ai,Harlander:2010cz,Harlander:2011fx,Buehler:2012cu,Marzani:2008az,Harlander:2009mq,Harlander:2009my,Pak:2009dg,Neumann:2014nha,Catani:2011qz,Campbell:2016yrh,Grazzini:2013bna,Grazzini:2015nwa,Campbell:2017aul,Gehrmann:2020oec,Cascioli:2014yka,Grazzini:2015hta,Heinrich:2017bvg,Kallweit:2018nyv,Gehrmann:2014fva,Grazzini:2016ctr,Grazzini:2016swo,Grazzini:2017ckn,deFlorian:2013jea,deFlorian:2016uhr,Grazzini:2018bsd,Baglio:2012np,Li:2016nrr,deFlorian:2019app} (see e.g.~\citere{Heinrich:2020ybq} for a review), and even first such 
calculations for $2 \to 3$ processes are emerging \cite{Chawdhry:2019bji,Kallweit:2020gcp,Czakon:2021mjy,Chawdhry:2021hkp}.}
triggered considerable advancements in the matching of NNLO QCD
corrections and parton
showers (NNLO+PS). The first method developed was the NNLOPS method based on
\minlo~\cite{Hamilton:2012rf,Hamilton:2013fea}, which achieves NNLO QCD
accuracy
through a reweighting of \minlo{} events. This method was
successfully employed for relatively simple processes, such as Higgs
production~\cite{Hamilton:2013fea}, Drell-Yan
production~\cite{Karlberg:2014qua} and associated Higgs
production~\cite{Astill:2016hpa,Astill:2018ivh}, i.e.\ to processes
that from a QCD point of view are just $2\to 1$ processes. The same
method was then also employed for $W^+W^-$ production, including the
decay of the $W$-bosons~\cite{Re:2018vac}. This paper showed
explicitly the limitations of the NNLOPS method, because in practice the multi-differential reweighting can not easily be
applied to more complicated processes, without making certain assumptions or approximations. 
About ten years 
ago, two more NNLO+PS methods were proposed: the {\sc UNNLOPS} one, which has only 
been applied to Higgs~\cite{Hoche:2014dla} and Drell-Yan
production~\cite{Hoeche:2014aia}, and the {\sc Geneva}
method~\cite{Alioli:2012fc,Alioli:2013hqa}. The latter was
subsequently modified, as far as the interface to the shower is
concerned, and applied to Drell-Yan \cite{Alioli:2015toa},
Higgsstrahlung \cite{Alioli:2019qzz}, photon pair
production~\cite{Alioli:2020qrd}, hadronic Higgs
decays~\cite{Alioli:2020fzf}, $ZZ$ production~\cite{Alioli:2021egp}, and $W\gamma$ production \cite{Cridge:2021hfr}.
Recently, the {\sc Geneva} method was reformulated using the transverse
momentum of the colour singlet rather than the jettiness variable and
applied to Drell-Yan~\cite{Alioli:2021qbf}.

Two years ago, the \minnlo{} method was
proposed~\cite{Monni:2019whf,Monni:2020nks}, whose underlying idea is
very similar to the \minlo approach that achieves NLO accuracy for
colour singlet plus zero and one jet simultaneously. The \minnlo
method exploits the close connection to transverse-momentum
resummation to include the relevant logarithmically enhanced and
constant terms to reach NNLO accuracy.  This method was first used to
reproduce known results for Higgs production and Drell-Yan~\cite{Monni:2019whf,Monni:2020nks} and it was applied more
recently to $Z\gamma$~\cite{Lombardi:2020wju} and $W^+W^-$
production~\cite{Lombardi:2021rvg}. Remarkably, although it was the 
last NNLO+PS method to appear, the \minnlo{} method
was the first to be extended and applied to the production of a 
coloured final state, namely top-quark pair production \cite{Mazzitelli:2020jio}.

In this work, we employ the \minnlo{} method to include NNLO QCD
corrections for $ZZ$ production in the \POWHEG{}
framework. Additionally, we present a NLO+PS \POWHEG{} calculation for
the loop-induced $gg \to ZZ$ process.  When combined, these
predictions, dubbed nNNLO+PS, become the most advanced theoretical
predictions for $ZZ$ production at the LHC, since they include the
highest perturbative accuracy in QCD available to date.  Spin
correlations, interferences and off-shell effects are included by
considering all contributions to the four-lepton final
state. Moreover, the matching to the parton shower renders it possible
to achieve a fully exclusive description at the level of hadronic
events.
In the future, the NNLO+PS predictions of our \minnlo{} $ZZ$ generator could be 
compared to those recently obtained in the {\sc Geneva} framework \cite{Alioli:2021egp}.

This manuscript is organized as follows: in \sct{sec:calculation} we
discuss in detail the calculation and implementation of the \minnlo
method for the $q\bar q$-initiated process and the \POWHEG{}
implementation for the loop-induced $gg$-initiated process.  We also
show how to avoid that the two-loop amplitudes, whose numerical
evaluation is very time-consuming, slow down our code in a
considerable way. Our phenomenological results for both cross sections
and distributions in $ZZ$ production are discussed in
\sct{sec:results}, where we present a comparison between showered,
fixed-order, and analytically resummed results at high accuracy for
various observables as well as a comparison of our nNNLO+PS
predictions to recent LHC data from CMS. We conclude in
\sct{sec:summary}.

\section{Outline of the calculation}
\label{sec:calculation}

\subsection{Description of the process}
\label{sec:process}

\begin{figure}[b]
  \begin{center}
\begin{subfigure}[b]{0.33\linewidth}
      \centering
      \resizebox{0.95\linewidth}{!}{
      \begin{tikzpicture}
        \begin{feynman}
          \vertex (a1) {\(q\)};
          \vertex[below=1.6cm of a1] (a2){\(\bar q\)};
          \vertex[right=2cm of a1] (a3);
          \vertex[right=2cm of a2] (a4);
          \vertex[right=1.5cm of a3] (a5);
          \vertex[right=1cm of a5] (a20);
          \vertex[above=0.3cm of a20] (a7){\(\ell^{\prime +}\)} ;
          \vertex[below=0.3cm of a20] (a8){\(\ell^{\prime -}\)};
          \vertex[right=1.5cm of a4] (a6);
          \vertex[right=1cm of a6] (a21);
          \vertex[above=0.3cm of a21] (a9){\(\ell^-\)} ;
          \vertex[below=0.3cm of a21] (a10){\(\ell^+\)};

          \diagram* {
            {[edges=fermion]
              (a1)--(a3)--[edge label'=\(q\)](a4)--(a2),
              (a7)--(a5)--(a8),
              (a10)--(a6)--(a9),
            },
            (a3) -- [boson, edge label=\(Z/\gamma^*\)] (a5),
            (a4) -- [boson, edge label=\(Z/\gamma^*\)] (a6),
          };

        \end{feynman}
      \end{tikzpicture}}
      \caption{}
        \label{subfig:zzres1}
    \end{subfigure}%
\begin{subfigure}[b]{0.33\linewidth}
      \centering
      \resizebox{0.95\linewidth}{!}{
\begin{tikzpicture}      
  \begin{feynman}
    \vertex (a1) {\(q\)};
    \vertex[below=1.6cm of a1] (a2){\(\bar q\)};
    \vertex[below=0.8cm of a1] (a3);
    \vertex[right=1.2cm of a3] (a4);

    \vertex[right=1cm of a4] (a5);

    \vertex[right=0.8cm of a5](a6);
    \vertex[above=0.3cm of a6](a7);

    \vertex[right=0.8cm of a7](a8);
    \vertex[above=0.3cm of a8](a9);

    \vertex[right=0.8cm of a9](a10);
    \vertex[above=0.2cm of a10](a11){\(\ell^{\prime -}\)};
    \vertex[below=0.2cm of a10](a12){\(\ell^{\prime +}\)};
    \vertex[below=0.7cm of a10](a13){\(\ell^-\)};
    \vertex[below=1.2cm of a10](a14){\(\ell^+\)};

     \diagram* {
       {[edges=fermion]
         (a1)--(a4)--(a2),
         (a14)--(a5)--[edge label=\(\ell^-\),inner sep=1.5pt,near end](a7)--(a13),
         (a12)--(a9)--(a11),
       },
       (a4) -- [boson, edge label=\(Z/\gamma^*\)] (a5),
       (a7) -- [boson, edge label=\(Z/\gamma^*\),inner sep=1.5pt,near end] (a9),
       };

  \end{feynman}
\end{tikzpicture}}
\vspace{0.4cm}
\caption{}
        \label{fig:zzres2}
\end{subfigure}
\begin{subfigure}[b]{0.33\linewidth}
      \centering
      \resizebox{0.95\linewidth}{!}{
      \begin{tikzpicture}
        \begin{feynman}
          \vertex (a1) {\(g\)};
          \vertex[below=1.6cm of a1] (a2){\(g\)};
          \vertex[right=1.2cm of a1] (a3);
          \vertex[right=1.2cm of a2] (a4);
          \vertex[right=1.5cm of a3] (a5);
          \vertex[right=1.2cm of a5] (a7);
          \vertex[right=1cm of a7] (a20);
          \vertex[above=0.3cm of a20] (a9){\(\ell^{\prime +}\)} ;
          \vertex[below=0.3cm of a20] (a11){\(\ell^{\prime -}\)};
          \vertex[right=1.5cm of a4] (a6);
          \vertex[right=1.2cm of a6] (a8);
          \vertex[right=1cm of a8] (a21);
          \vertex[above=0.3cm of a21] (a10){\(\ell^-\)} ;
          \vertex[below=0.3cm of a21] (a12){\(\ell^+\)};

          \diagram* {
             {[edges=gluon]
              (a1)--(a3),
              (a2)--(a4),
            },
            {[edges=fermion]
              (a3)--(a5)--(a6)--(a4)--[edge label=\(q\)](a3),
              (a9)--(a7)--(a11),
              (a12)--(a8)--(a10),
            },
            (a5) -- [boson, edge label=\(Z/\gamma^*\)] (a7),
            (a6) -- [boson, edge label=\(Z/\gamma^*\)] (a8),
          };

        \end{feynman}
      \end{tikzpicture}}
\caption{}
        \label{fig:zzres3}
\end{subfigure}
\end{center}
\vspace{-0.3cm}
  \caption{\label{DiagramsZZ} Sample Feynman diagrams for $ZZ$
    production with four charged leptons in the final state. Panels
    (a) and (b): tree-level diagrams of the quark annihilation
    ($q\bar{q}$) channel; \mbox{Panel (c)}: loop-induced diagram in
    the gluon fusion ($gg$) channel.  }
\label{fig:zzres}  
\end{figure}
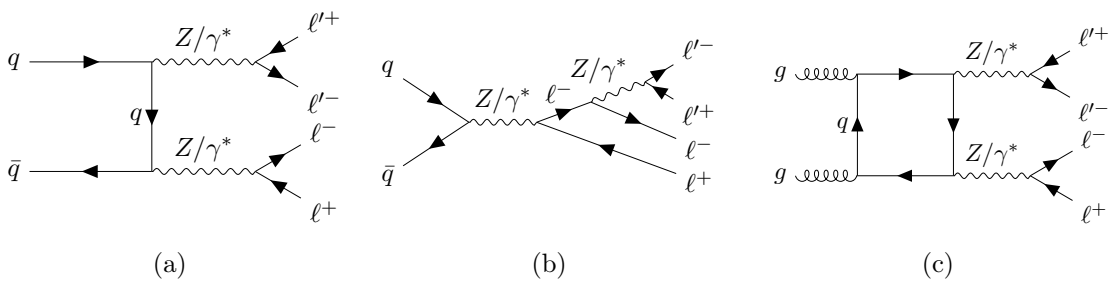

We study the process
\begin{align}
  pp \to \ell^+\ell^-\ell^{(\prime)+}\ell^{(\prime)-}
\label{eq:process}
\end{align}
for any combination of charged leptons $\ell,\ell^\prime
\in\{e,\mu,\tau\}$.  While at the matrix-element level our calculation
is based on the different-flavour channel $\ell\neq\ell^\prime$, at
the event-generation level arbitrary combinations of charged leptons
can be considered, both with different flavours $\ell\neq\ell^\prime$
and with same flavours $\ell=\ell^\prime$ (in the latter case
interference effects when exchanging the charged leptons, which are
typically at the 1-2\% level~\cite{Grazzini:2015hta}, are neglected).
Moreover, lepton masses are included via reshuffling of the momenta
in the event generation.  For simplicity and without loss of
generality we consider only the process \mbox{$p p \to e^+e^-
  \mu^+\mu^-$} here, which we will refer to as
\zz{} production in the following.  By including all resonant and
non-resonant topologies leading to this process, off-shell effects,
interferences and spin correlations are taken into account.
Sample diagrams are shown in \fig{fig:zzres} and they include:
\begin{itemize}
	\item [(a)]  tree-level double-resonant t-channel $ZZ$ production in the $q\bar{q}$ channel;
	\item [(b)]  tree-level single-resonant s-channel DY topologies in the $q\bar{q}$ channel;
	\item [(c)]  loop-induced $ZZ$ production in the $gg$ channel.
\end{itemize}
The loop-induced contribution proceeds through a quark loop and enters
the cross section at $\mathcal{O}(\as^2)$, i.e. it is part of the NNLO
QCD corrections.  Since this contribution is enhanced by the large
gluon-gluon luminosity at LHC energies, it yields a relatively large
fraction of the NNLO corrections
\cite{Gehrmann:2014fva,Grazzini:2016ctr}. Moreover, it is known that
at NLO QCD \cite{Caola:2015psa,Grazzini:2018owa},
i.e.\ $\mathcal{O}(\as^3)$, its relative correction is very sizable
(about a factor of two), which renders it the most significant
contribution to \zz{} production at $\mathcal{O}(\as^3)$.

We include the most accurate currently available information in QCD
perturbation theory for both the $q\bar{q}$-initiated and the
$gg$-initiated process, and match them consistently with a parton
shower. Thus, we calculate NNLO+PS predictions in the $q\bar{q}$
channel by means of the \minnlo{} method
\cite{Monni:2019whf,Monni:2020nks,Lombardi:2020wju} and NLO+PS
predictions in the $gg$ channel using the \POWHEG{} approach
\cite{Nason:2004rx,Frixione:2007vw,Alioli:2010xd}. Our ensuing result
is dubbed as nNNLO+PS, as the NLO corrections to the loop-induced
gluon-fusion contribution are of $\mathcal{O}(\as^3)$. These
corrections are separately gauge-invariant and constitute the most
significant N$^3$LO corrections, as pointed out before.

\subsection[\minnlo{} for $q\bar{q}\to ZZ$ production]{\minnlo{} for \boldmath{$q\bar{q}\to ZZ$} production}
\label{sec:minnlo}

In this section we present the implementation of a NNLO+PS generator
for $ZZ$ production in the $q\bar{q}$ channel by means of the
\minnlo{} method.
We first sketch the \minnlo{} method by introducing its essential
ingredients and then discuss its practical implementation within the
\POWHEGBOXRES{} framework \cite{Jezo:2015aia} and linking \Matrix{}
\cite{Grazzini:2017mhc}.

\subsubsection{The \minnlo{} method for colour-singlet production}

\minnlo has been formulated in~\citere{Monni:2019whf}, optimised for
$2\rightarrow1$ processes in \citere{Monni:2020nks} and later extended
to generic colour-singlet processes in \citere{Lombardi:2020wju} and
to heavy-quark pair production in \citere{Mazzitelli:2020jio}.  We
refer to those publications for a detailed description of the
\minnlo{} method and here we only sketch the procedure adopting a
simplified notation.

The \minnlo method includes NNLO corrections in the event generation
of a system \F{} of colour-singlet particles.  It involves essentially
three steps: in the first one (\stepone{}) \F{} is generated in
association with one light parton at NLO according to the \POWHEG{}
method~\cite{Nason:2004rx,Nason:2006hfa,Frixione:2007vw,Alioli:2010xd},
inclusively over the radiation of a second light parton. The second
step (\steptwo{}) characterizes the \minnlo{} approach, and it
corrects the limit in which the light partons become unresolved by
supplementing the appropriate Sudakov form factor and higher-order
terms, such that the simulation remains finite as well as NNLO
accurate for inclusive \F{} production. In the third step
(\stepthree{}), the kinematics of the second radiated parton
(accounted for inclusively in \stepone{}) is generated through the
\POWHEG{} method to preserve the NLO accuracy of the \F{}+$1$-jet
cross section, and subsequent radiation is included through the parton
shower. In these three steps all emissions are appropriately ordered
(when using a $p_T$-ordered shower) and the applied Sudakov matches
the leading logarithms resummed by the parton shower.  Thus, the
\minnlo{} approach preserves the (leading logarithmic) accuracy of the
parton shower.

The fully differential \minnlo{} cross section can be expressed
through the \POWHEG{} formula for the production of a colour singlet
plus one light parton (\FJ{}) with a modified content of the $\bar B$
function:
\begin{align}
{\rm d}\sigma_{\rm\scriptscriptstyle F}^{\rm MiNNLO_{PS}}=\bar{B}^{\,\rm MiNNLO_{\rm PS}}\,\times\,\left\{\Delta_{\rm pwg}(\Lambda_{\rm pwg})+\int {\rm d}\Phi_{\rm rad}\Delta_{\rm pwg}(p_{T,{\rm rad}})\,\frac{R_{\scriptscriptstyle\rm FJ}}{B_{\scriptscriptstyle\rm FJ}}\right\}\,,
\end{align}
where $\Delta_{\rm pwg}$ is the \POWHEG{} Sudakov form factor,
$\Phi_{\tmop{rad}} $ and $\ptrad$ are the phase space and the
transverse momentum of the second radiation, respectively, and
$B_{\scriptscriptstyle\rm FJ}$ and $R_{\scriptscriptstyle\rm FJ}$
denote the squared tree-level matrix elements for \FJ{} and \FJJ{}
production, respectively.
The content of the curly brackets generates the second QCD emission
according to the \POWHEG{} method, as described in \stepthree{} above,
with a default cutoff of $\LambdaPWG=0.89$\,GeV.  The $\bar{B}$
function contains the same contributions to generate the first
emission (and inclusively the second emission) as in a standard \FJ{}
\POWHEG{} calculation, described in \stepone{} above, but it is
modified according to the \minnlo{} procedure in order to reach NNLO
accuracy for inclusive \F{} production, as discussed in \steptwo{}
above.  Symbolically, it can be written as
\begin{align}
\label{eq:minnlo}
\bar{B}^{\,\rm MiNNLO_{\rm PS}}\sim e^{-S}\,\Big\{{\rm d}\sigma^{(1)}_{\scriptscriptstyle\rm FJ}\big(1+S^{(1)}\big)+{\rm d}\sigma^{(2)}_{\scriptscriptstyle\rm FJ}+\left(D-D^{(1)}-D^{(2)}\right)\times F^{\rm corr}\Big\}\,,
\end{align}
with ${\rm d}\sigma^{(1,2)}_{\scriptscriptstyle\rm FJ}$ being the
first- and second-order contribution to the differential \FJ{} cross
section and $e^{-S}$ denoting the Sudakov form factor for the
transverse momentum of \F{}. Note that in the \minnlo{} approach the
renormalization and factorization scales are evaluated as
$\muR\sim\muF\sim \pt$, where $\pt{}$ is the transverse momentum of
\F{}.  The third term in \eqn{eq:minnlo} is of order
$\as^3(p_{\text{\scalefont{0.77}T}})$ and it adds the relevant
(singular) contributions so that the integration over $\pt$ yields a
NNLO accurate result~\cite{Monni:2019whf}. Regular contributions at
this order are of subleading nature.  The function $D$ is derived from
the transverse-momentum resummation formula, which can be expressed
fully differentially in the Born phase space of \F{} as
\begin{align}
\label{eq:resum}
{\rm d}\sigma_{\scriptscriptstyle\rm F}^{\rm res}=\frac{{\rm d}}{{\rm d}\pt}\left\{e^{-S}\mathcal{L}\right\}=e^{-S}\underbrace{\left\{-S^\prime\mathcal{L}+\mathcal{L}^\prime\right\}}_{\equiv D}\,,
\end{align}
with $\mathcal{L}$ being the luminosity factor up to NNLO that
includes the squared hard-virtual matrix elements for \F{} production
and the convolution of the collinear coefficient functions with the
parton distribution functions (PDFs).  In fact, \eqn{eq:minnlo}
follows directly from matching \eqn{eq:resum} with the fixed-order
cross section ${\rm d}\sigma_{\scriptscriptstyle\rm FJ}$, when using a
matching scheme where the Sudakov form factor is factored out, i.e.\
\begin{align}
 & {\rm d}\sigma_{\scriptscriptstyle\rm F}^{\rm res}+[{\rm d}\sigma_{\scriptscriptstyle\rm FJ}]_{\rm f.o.}-[{\rm d}\sigma_{\scriptscriptstyle\rm F}^{\rm res}]_{\rm f.o.}=e^{-S}\bigg\{D+[{\rm d}\sigma_{\scriptscriptstyle\rm FJ}]_{\rm f.o.}\,\underbrace{\frac{1}{[e^{-S}]_{\rm f.o.}\,}}_{1+S^{(1)}\cdots}\underbrace{\,\,-\,\frac{[{\rm d}\sigma_{\scriptscriptstyle\rm F}^{\rm res}]_{\rm f.o.}\,\,}{[e^{-S}]_{\rm f.o.}}}_{-D^{(1)}-D^{(2)}\cdots}\bigg\}\\
&\quad\quad\quad\quad\quad\quad\quad\,\,=e^{-S}\,\bigg\{{\rm d}\sigma^{(1)}_{\scriptscriptstyle\rm FJ}\big(1+S^{(1)}\big)+{\rm d}\sigma^{(2)}_{\scriptscriptstyle\rm FJ}+\left(D-D^{(1)}-D^{(2)}\right)+\mathcal{O}(\as^4)\bigg\}\,,\nonumber
\end{align}
where $[\cdots]_{\rm f.o.}$ denotes the expansion up to a given fixed
order in $\as$.  Finally, $F^{\rm corr}$ in \eqn{eq:minnlo} determines
the appropriate function to spread the NNLO corrections in the \FJ{}
phase space, which is necessary to include those corrections in the
context of an \FJ{} \POWHEG{} calculation. Note also that one could
either truncate the third term in \eqn{eq:minnlo} at the third order
$\left(D-D^{(1)}-D^{(2)}\right)=D^{(3)}+\mathcal{O}(\as^4)$
\cite{Monni:2019whf}, or keep the terms of $\mathcal{O}(\as^4)$ and
higher \cite{Monni:2020nks}, which are beyond accuracy, in order to
preserve the total derivative in \eqn{eq:resum}. We employ the latter
option here.

\subsubsection{Practical implementation in \POWHEGBOXRES+\Matrix}
\label{sec:practical}

In the following we provide some information on our implementation of
a \minnlo{} generator for $ZZ$ production in the $q\bar{q}$ channel
within the \POWHEGBOXRES{} framework~\cite{Jezo:2015aia}. Our NLO+PS
generator for the loop-induced $gg$ channel is discussed in the next
section.  We stress that, while we distinguish these processes as
$q\bar q$ and $gg$, in their higher-order corrections of course all
the relevant partonic initial states are consistently included.

Since no implementation for $ZZ$+jet production was available in
\POWHEGBOX{} to date, the first step was to implement this process in
the \POWHEGBOXRES{} framework.  We have implemented all relevant
flavour channels and, in addition, adapted the routine {\tt
  build\char`_resonance\char`_histories} of \POWHEGBOXRES{} such that
it is capable of automatically constructing the correct resonance
histories. The tree-level single and double real matrix elements
for $e^+ e^- \mu^+ \mu^-$+$1,2$-jet production and the one-loop
amplitude for $e^+ e^- \mu^+ \mu^-$+1-jet production are evaluated
through
\OpenLoops{}~\cite{Cascioli:2011va,Buccioni:2017yxi,Buccioni:2019sur}.

In a second step, we have employed the \minnlo{} method to obtain
NNLO+PS predictions for $ZZ$ production in the $q\bar{q}$ channel.  In
particular, we made use of the implementation of the \minnlo{} method
that was developed and applied to $Z\gamma$ production in
\citere{Lombardi:2020wju}. The respective tree-level and one-loop
$q\bar{q}\to e^+ e^- \mu^+ \mu^-$ amplitudes are also evaluated
through \OpenLoops{}, while the two-loop helicity amplitudes have been
obtained by extending the interface to \Matrix{}\,\cite{Grazzini:2017mhc} 
developed in \citere{Lombardi:2020wju} to $ZZ$ production. The
evaluation of the two-loop coefficients in this implementation relies
on the code \textsc{VVamp}\,\cite{hepforge:VVamp} and is based on the
calculation of \citere{Gehrmann:2015ora}.

As discussed in \citere{Lombardi:2021rvg} for \ww{} production, the
evaluation of the two-loop helicity amplitudes for massive diboson
processes is particularly demanding from a computational point of
view.  In \citere{Lombardi:2021rvg} this problem was circumvented by
constructing a set of interpolation grids for the two-loop
coefficients that achieves their fast on-the-fly evaluation.  In this
work we pursue a different strategy: we exploit the possibility of
reweighting the events at the generation level (i.e. stage 4 in
\POWHEGBOX{}) to include the two-loop contribution. In combination
with a suitable caching system of the two-loop amplitude that we
implemented this allows us to omit the evaluation of the two-loop
contribution entirely in the calculation up to stage 4, where it needs
to be evaluated only once per event.\footnote{Note that in order for
  the caching to work properly and not having to reevaluate the
  two-loop amplitude for every scale variation in the event
  reweighting, we have set the parameter {\tt
    rwl\char`_group\char`_events\,1} in the input file, which ensures
  that the events are reweighted one-by-one instead of in batches.}
To be more precise, we have implemented a new flag
(\texttt{run\char`_mode}), which is accessible from the \POWHEG{}
input file, and allows the user to switch between four different ways
of running the code.
Either the full calculation is performed including the two-loop contributions throughout ({\tt run\char`_mode\,1}), or one completely drops the NNLO corrections provided by \minnlo{}, specifically the terms $\left(D-D^{(1)}-D^{(2)}\right)\times F^{\rm corr}$
in \eqn{eq:minnlo}, 
thus effectively reproducing \minlo{} predictions ({\tt run\char`_mode\,2}).
Alternatively, the evaluation of two-loop amplitude
can be omitted only in the grid setup,
i.e. stage 1 in \POWHEGBOX{}, ({\tt run\char`_mode\,3}), or completely
({\tt run\char`_mode\,4}).  For all results presented in this
manuscript we run the code with the last option 
\texttt{run\char`_mode\,4}, i.e.\ without evaluating the 
computationally expensive two-loop
amplitude.  In this way, also the generation of the events is faster.
However, once an event has been accepted, it is reweighted such that
the two-loop contribution is included by resetting the
\texttt{run\char`_mode} keyword in the event reweight information of
the \POWHEG{} input file.  As a result the two-loop amplitude is
evaluated only once for each event, considerably improving the
efficiency of the code.  Moreover, following the same logic we can
also compute \minlo weights in parallel to the generation of \minnlo
ones using the appropriate setting for \texttt{run\char`_mode} in the
event reweight information.  We have first validated our
implementation in an inclusive setup, requiring only a suitable
$Z$-mass window for the opposite-charge same-flavour dilepton pairs.
Here we compared the inclusive cross section at the Les Houches event
(LHE) level obtained at stage 4 with the one computed at stage 2 when
including the two-loop contribution, finding excellent agreement.
Another very robust cross-check of the reweighting procedure is
provided by the comparison of the \minlo{} results obtained directly
or through reweighting, which also agree perfectly.

Our calculation involves the evaluation of several convolutions with
the PDFs, for which we employ \noun{hoppet}~\cite{Salam:2008qg}.  The
evaluation of the polylogarithms entering the collinear coefficient
functions is done through the \noun{hplog}
package~\cite{Gehrmann:2001pz}.

Finally, let us summarize some of the most relevant (non-standard)
settings that we have used to produce NNLO+PS accurate \zz{} events in
the $q\bar{q}$ channel.  For more detailed information on those
settings we refer to \citeres{Monni:2020nks,Mazzitelli:2020jio}.  To avoid spurious
contributions from higher-order logarithmic terms at large $\pt$, we
make use of a modified logarithm that is defined such that it smoothly 
vanishes at $\pt$ equal to or larger than the invariant mass of the system, as used in 
\citere{Mazzitelli:2020jio}.  As far as the renormalization
and factorization scales are concerned, we use the standard \minnlo{}
scale setting in eq.\,(14) of \citere{Monni:2020nks} at small $\pt$,
while in the NLO \zz{}+jet cross section at large $\pt$ the scale
setting is changed to the one in eq.\,(19) of \citere{Monni:2020nks}
by activating the option {\tt largeptscales\,1}.  Since those scale
settings have been defined with $\Q =0$\,GeV, the Landau singularity
is regulated by freezing the strong coupling and the PDFs for scales
below $0.8$\,GeV.  Finally, as recommended for processes with jets in
the final state, we turn on the option \texttt{doublefsr\,1} of the
\POWHEGBOX{}, see \citere{Nason:2013uba} for details.  For the
parton-shower we have used the standard settings, also for the recoil
scheme (namely a global recoil scheme for initial state radiation,
with \texttt{SpaceShower:dipoleRecoil\,0}).

\subsection[NLO+PS for $gg\to ZZ$ production]{NLO+PS for \boldmath{$gg\to ZZ$} production}
\label{sec:gg}
As discussed before, the leading-order contribution to the
loop-induced gluon fusion process enters the \zz{} cross section at
$\mathcal{O}(\as^2)$.  Thus, it constitutes a NNLO correction relative
to the LO part of the $q\bar{q}$ channel, but it is significantly
enhanced by the large gluon-gluon luminosities. It is therefore
mandatory to include also the NLO corrections to the loop-induced
gluon fusion contribution in any precision study of $ZZ$ production
that compares theory and data.

We have implemented a NLO+PS generator for loop-induced $ZZ$
production in the $gg$ channel within the \POWHEGBOXRES{} framework.
Note that in addition to continuum $ZZ$ production as shown in
\fig{DiagramsZZ}\,(c) we also include the contribution mediated by a
Higgs boson.  The calculation of these loop-induced processes is
effectively of similar complexity as a NNLO calculation, as far as the
amplitude evaluation is concerned. Despite that, the matching to the
parton shower through the \POWHEG{} method, which is essentially
automated in \POWHEGBOXRES{}, can be applied to loop-induced processes
as well, without any further conceptual issues.  However such an NLO
calculation requires the evaluation of both one-loop and two-loop
helicity amplitudes and the process at hand is numerically
substantially more demanding than a tree-level one, since the
evaluation time of the one-loop and two-loop amplitudes is much slower
and the stability of the one-loop matrix elements with one emitted
real parton is challenged in the infrared regions. To cope with these
numerical issues, we have implemented and exploited a number of
handles within \POWHEGBOXRES{}, which will be discussed below.

For the implementation in \POWHEGBOXRES{}, we have specified the
relevant flavour channels and hard-coded also the resonance channels
of the process, as the automatic determination of the latter via the
already mentioned routine {\tt build\char`_resonance\char`_histories}
is not available yet for loop-induced processes. At NLO, all relevant
partonic channels, namely $gg$, $gq$, $qg$, $q\bar{q}$ and the
charge-conjugated ones, are included.  To unambiguously define the NLO
corrections to the loop-induced gluon-fusion process for each of those
initial states, we follow the approach introduced in
\citere{Grazzini:2018owa} and include all diagrams that involve a
closed fermion loop with at least one vector boson attached.  The
one-loop amplitudes with zero and one jet are evaluated through
\OpenLoops{}~\cite{Cascioli:2011va,Buccioni:2017yxi,Buccioni:2019sur}.
For this purpose, we have adapted the \OpenLoops{} interface in
\POWHEGBOXRES{} developed in \citere{Jezo:2016ujg} to deal with
loop-induced processes. As a cross-check, we have also interfaced
\Recola{} to \POWHEGBOXRES{} and found full agreement for all one-loop
amplitudes.  For the two-loop helicity $gg\to
\ell^+\ell^-\ell^{(\prime)+}\ell^{(\prime)-}$ amplitudes we exploit
their implementation within \Matrix{}~\cite{Grazzini:2017mhc}, which
is based on the evaluation of the two-loop coefficients through
\textsc{VVamp}~\cite{Gehrmann:2015ora} from their calculation in
\citere{vonManteuffel:2015msa}.  To this end, we have extended the
interface of \POWHEGBOXRES{} to \Matrix{} developed in
\citere{Lombardi:2020wju} to include the $gg\to
\ell^+\ell^-\ell^{(\prime)+}\ell^{(\prime)-}$ two-loop
amplitudes. Also here the evaluation of the two-loop coefficients
through \textsc{VVamp} is very slow, lasting from a few seconds to
several tens of seconds. Since this leads to a severe bottleneck in
the calculation and especially in the event generation, we have
implemented a caching system for the two-loop corrections and we
include them only through event reweighting. This is very similar in
spirit to the way the two-loop hard function is included in the
\minnlo{} generator in the $q\bar{q}$ channel, as described in the
previous section.  Our calculation includes the full top-quark mass
effects, except for the two-loop $gg\to
\ell^+\ell^-\ell^{(\prime)+}\ell^{(\prime)-}$ amplitudes, where they
are not known to date.\footnote{For the case of on-shell $gg\to ZZ$
  production the full top-quark mass dependence was recently
  calculated in \citeres{Agarwal:2020dye,Bronnum-Hansen:2021olh}.}
Instead, we follow the same approach as \citere{Grazzini:2018owa} and
include them approximately through a reweighting of the massless
two-loop amplitude with the ratio of the one-loop result including
massive loops to the one with only massless loops. Since here we are
interested in the $ZZ$ signal region, such reweighting is expected to
work extremely well. In fact, \citere{Alioli:2021wpn} recently
confirmed that using an asymptotic expansion in the top-quark mass
leads to practically identical results as using such reweighting, as
long as one sticks to the validity range of the expansion
itself.

\begin{figure*}[t]
\includegraphics[width=0.49\textwidth,height=7.85cm,keepaspectratio]{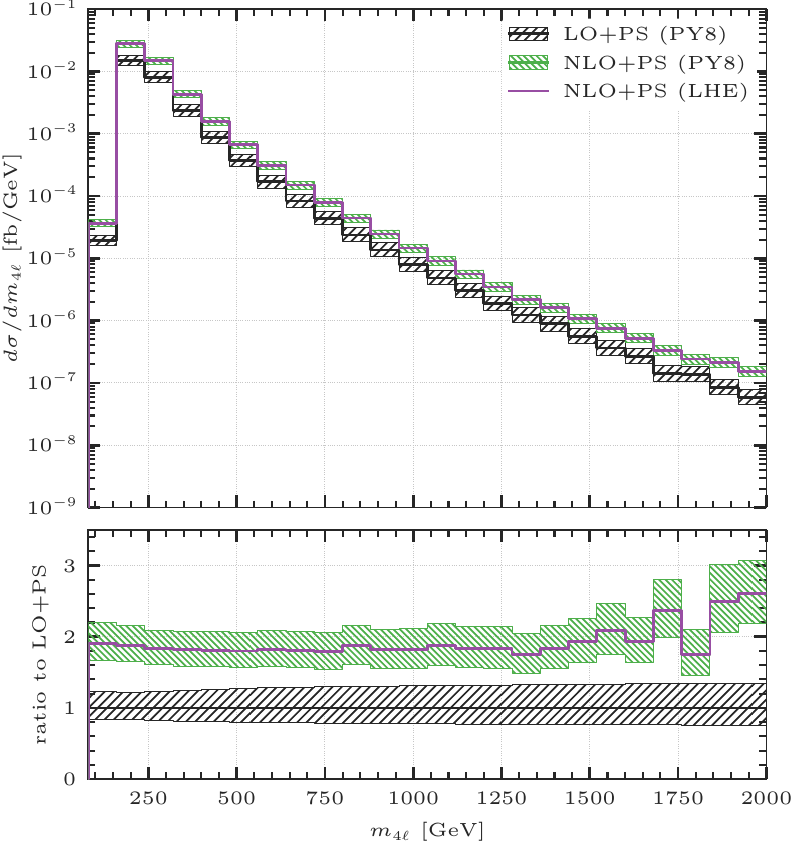}%
\hspace{2.5mm}
\includegraphics[width=0.49\textwidth,height=7.85cm,keepaspectratio]{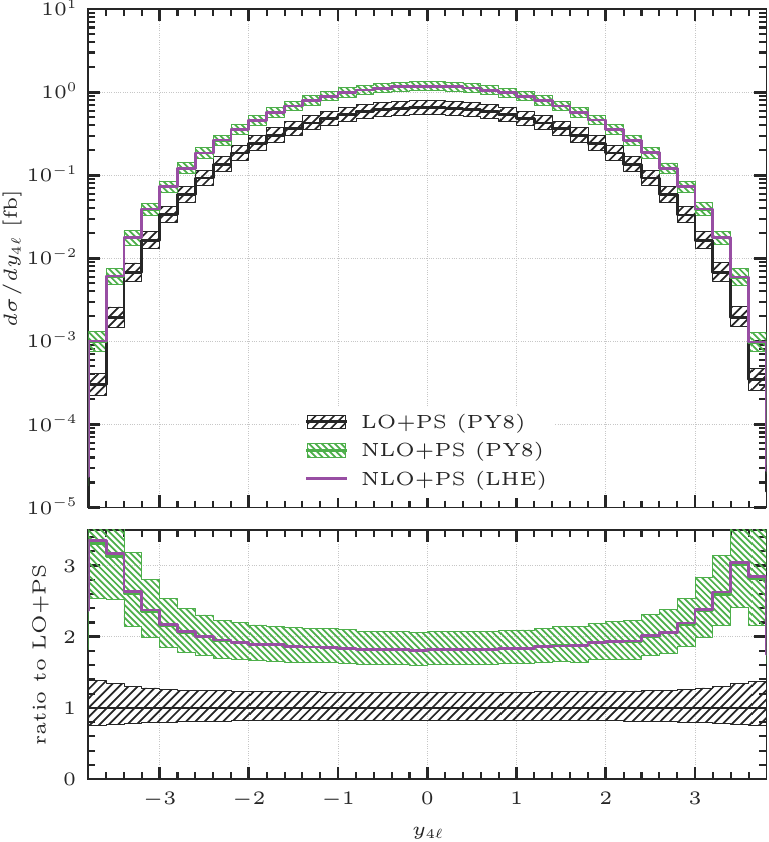}%
\\
\\
\includegraphics[width=0.49\textwidth,height=7.85cm,keepaspectratio]{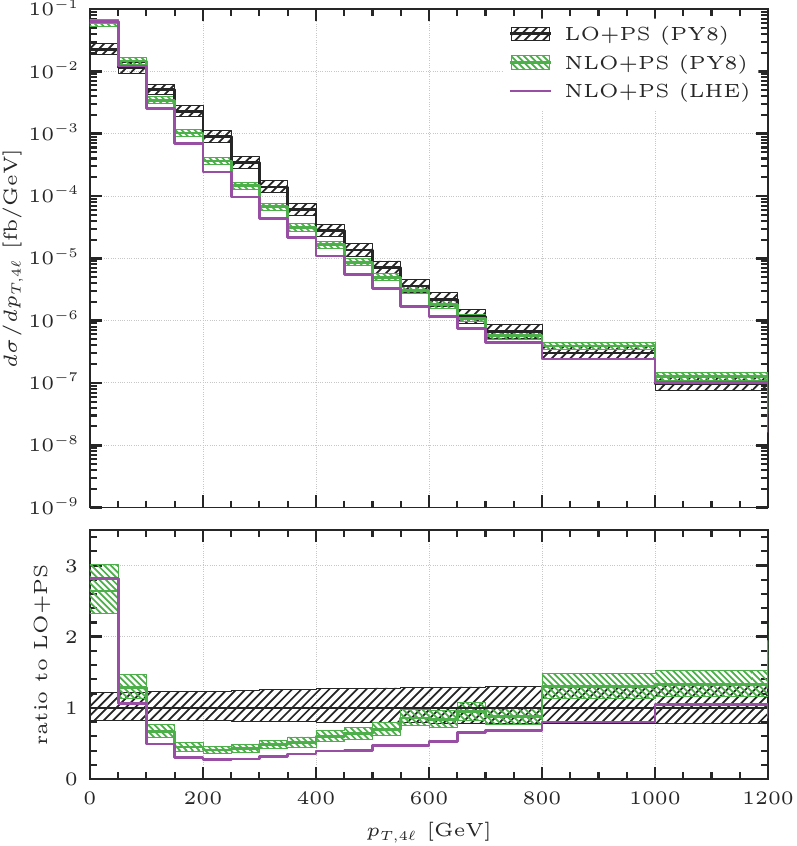}%
\hfill
\includegraphics[width=0.49\textwidth,height=7.85cm,keepaspectratio]{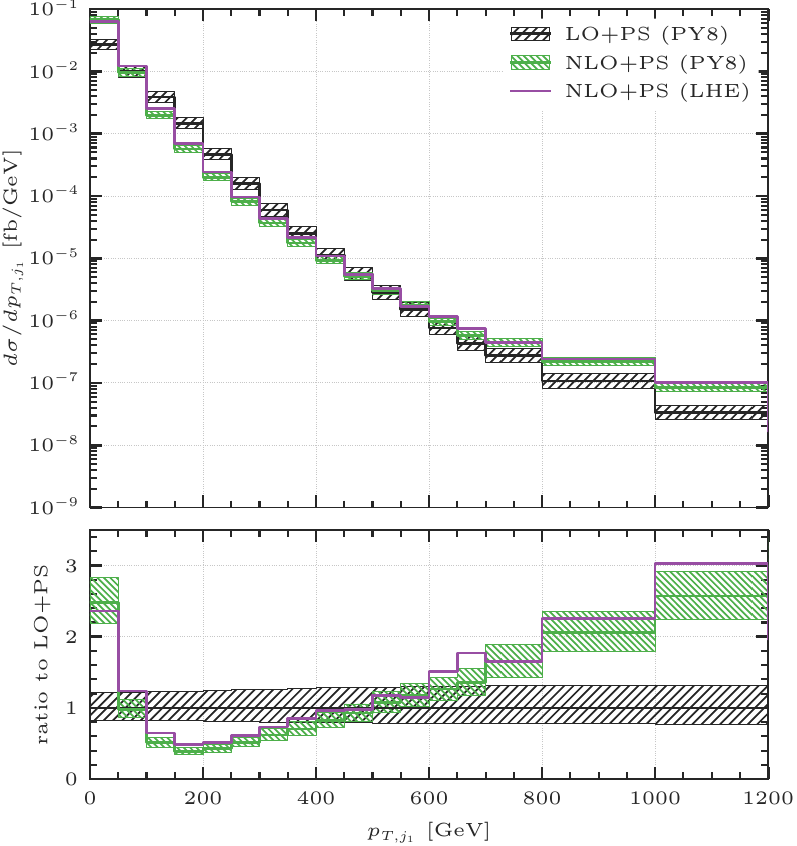}%
\hspace{2mm}
  \caption{Predictions for $ZZ$ production in the loop-induced $gg$
    channel at LO+PS and NLO+PS. For reference also the LHE-level
    central result at NLO is plotted. Shown are the distributions in
    the invariant mass, rapidity and transverse momentum of the
    four-lepton system, and in the transverse momentum of the leading
    jet.}
	\label{fig:gg}
\end{figure*}

Very recently, \citere{Alioli:2021wpn} presented a completely
independent implementation of a NLO+PS generator for loop-induced $ZZ$
production in the $gg$ channel within the \POWHEGBOXRES{}
framework. We have compared our calculation to theirs both at the
level of individual phase-space points and of the integrated cross
sections, and we have found perfect agreement when applying the same
approximation for the two-loop virtual corrections.\footnote{We would
  like to thank the authors of \citere{Alioli:2021wpn} for providing
  the {\tt gg4l} code and in particular Jonas Lindert for very helpful
  correspondence.}  Since, although developed independently, the two
calculations are essentially interchangeable (both developed in
\POWHEGBOXRES{} using \OpenLoops{} and \textsc{VVamp}), we advocate
that it is equivalent to use either code and combine the results
subsequently with our \minnlo{} generator in the $q\bar{q}$ channel to
obtain nNNLO+PS accurate results.

To better control the numerical stability of the calculation we have
implemented settings similar to those reported in
\citere{Alioli:2021wpn}: in particular we apply small (0.5\,GeV)
generation cuts on the transverse momentum of the four-lepton system
and of each $Z$ boson.  Moreover, we exploit the stability system of
\OpenLoops{} and set the parameter {\tt stability\char`_kill 0.01} to
remove the remaining unstable points. We have further modified the
code in such a way that, whenever the real-emission contribution is
set to zero by one of the previous stability checks, also the
respective counter terms are set to zero. Finally, we use {\tt
  withdamp 0} in order not to split the real cross section into a
singular and a remnant contribution as the considerably small value of
the latter leads to numerical issues when generating events. The same
is true for the regular contribution that contains only the $q\bar{q}$
channel: being completely negligible, we have turned it off
for all results obtained in this paper.

Since in the upcoming section we study phenomenological results for
the full $p p \rightarrow e^+ e^- \mu^+ \mu^-$ process, we show some
plots for the loop-induced $gg$ channel separately in \fig{fig:gg},
both at LO and at NLO.  The settings and inputs that we use here
correspond to those introduced in \sct{sec:setup} in the inclusive
setup (\setupinclusive{}) with just a $Z$-mass window applied between
$60$\,GeV and $120$\,GeV. The renormalization and factorization scales
are set to $\muR=\muF=\sqrt{m_{4\ell}^2+p_{T,4\ell}^2}$, where
$m_{4\ell}$ and $p_{T,4\ell}$ are the invariant mass and the
transverse momentum of the four-lepton system, respectively. The
uncertainty bands are obtained through a standard seven-point scale
variation.  For the genuine NLO-accurate quantities shown in
\fig{fig:gg}, namely $m_{4\ell}$ and the rapidity of the four-lepton
system ($y_{4\ell}$), we find results that are completely in line with
the findings of previous fixed-order calculations
\cite{Caola:2015psa,Grazzini:2018owa}, which is expected since shower
effects are negligible for those observables, as one can 
see from the LHE results. In particular, NLO
corrections are sizable and increase the value of the inclusive cross
section by almost a factor of two, with scale uncertainties at
the level of 10-15\%. In certain phase-space regions, like in the tail
of the $m_{4\ell}$ distribution, the NLO corrections can even become
significantly larger than a factor of two. However, in those regions
the relative impact of the loop-induced $gg$ contribution is reduced.
When looking at the transverse-momentum spectrum of the four-lepton
system ($p_{T,4\ell}$) and of the leading jet ($p_{T,j_1}$) in
\fig{fig:gg}, the importance of matching to the parton shower becomes
clear: at LO only the parton shower fills those distributions and at
NLO it still provides a substantial correction. In fact, in a
fixed-order calculation both observables would diverge, and therefore
be unphysical, at small transverse momenta. It is interesting to
notice that the LO+PS result is actually higher than the NLO+PS one in
the intermediate $p_{T,4\ell}$ region before it falls off
steeply. This region is completely filled by the shower, whose
starting scale by default is set to $m_{4\ell}$ in the LO
calculation. The fact that $m_{4\ell}$ is on average relatively large
explains why the shower fills the spectrum even at such high
transverse momenta.

\section{Phenomenological results}
\label{sec:results}
In this section we present phenomenological results for the process $p
p \rightarrow e^+ e^- \mu^+ \mu^-$.
After discussing our setup in \sct{sec:setup}, we compare our \minnlo
predictions for integrated cross sections (\sct{sec:minnlo-int}) and
at the differential level (\sct{sec:minnlo-diff}) against fixed-order
predictions at NNLO accuracy, \minlo results and experimental data
from the CMS experiment~\cite{CMS:2020gtj}.

\subsection{Input parameters and setup}
\label{sec:setup}

We consider proton--proton collisions at the LHC with a center-of-mass
energy of 13\,TeV and present predictions for $p p \rightarrow e^+ e^-
\mu^+ \mu^-$ production.
We use the complex-mass scheme~\cite{Denner:1999gp} throughout and set
the electroweak (EW) inputs to their PDG~\cite{Zyla:2020zbs} values:
$G_F = 1.16639 \times 10^{-5}$~GeV$^{-2}$, $m_W = 80.385$~GeV,
$\Gamma_W = 2.0854$~GeV, $m_Z = 91.1876$~GeV, $\Gamma_Z = 2.4952$~GeV,
$m_H = 125$~GeV and $\Gamma_H = 0.00407$~GeV. We set the on-shell
top-quark mass to $m_t = 173.2$~GeV, and $\Gamma_t =1.347878$~GeV is
used.
We determine the other EW parameters in the $G_\mu$ scheme with the EW
coupling $\alpha_{G_\mu} = \sqrt{2}/\pi G_\mu |(m_W^2 - i \Gamma_W
m_W) \sin^2 \theta_W| $ and the EW mixing angle $\cos^2 \theta_W =
(m_W^2 - i \Gamma_W m_W) / (m^2_Z - i \Gamma_Z m_Z)$.  We use the
NNPDF3.1~\cite{Ball:2017nwa} NNLO set with $\as=0.118$ via the
\textsc{lhapdf} interface~\cite{Buckley:2014ana} for all our
predictions.  For \minlo{} and \minnlo{}, the PDFs are read by
\textsc{lhapdf} and evolved internally by
\textsc{hoppet}~\cite{Salam:2008qg} as described in
\citere{Monni:2019whf}.  The central factorization and renormalization
scales are set as discussed in \sct{sec:practical} for the \minnlo{}
$ZZ$ generator in the $q\bar{q}$ channel and as given in \sct{sec:gg}
for the loop-induced $gg$ channel.  Scale uncertainties are estimated
by varying $\muF$ and $\muR$ around their central value by a factor of
two in each direction, while keeping the minimal and maximal values
with the constraint $0.5 \leq \muR /\muF \leq 2$.

\renewcommand{\baselinestretch}{1.5}
\begin{table}[b]
\centering
  \begin{tabular}{l|cc}
     & \setupinclusive & \setupfiducial \\
    \toprule
              $Z$-mass window & $60$\,GeV$ < m_{Z_1},m_{Z_2} < 120$\,GeV & $60$\,GeV$ < m_{Z_1},m_{Z_2}< 120$\,GeV\\[0.1cm]
    lepton cuts & $m_{\ell^+\ell^-} > 4 \, {\rm GeV}$  & $\begin{array}{c}
   p_{T,\ell_1} > 20\,{\rm GeV}, \quad p_{T,\ell_2} > 10 \, {\rm GeV},  \\[-0.15cm]
   p_{T,\ell_{3,4}} > 5 \, {\rm GeV}, \quad |\eta_\ell | < 2.5,\\[-0.15cm]
   m_{\ell^+\ell^-} > 4 \, {\rm GeV}  \\    
 \end{array}$
 \end{tabular}
 \renewcommand{\baselinestretch}{1.0}
  \caption{Inclusive and fiducial cuts used to the define the \setupinclusive and \setupfiducial phase space regions~\cite{CMS:2020gtj}. See text for more details.}
   \label{tab:cuts}
\end{table}
 \renewcommand{\baselinestretch}{1.0}

By combining the \minnlo{} $q\bar{q}$ results and loop-induced $gg$
results at (N)LO+PS, we obtain predictions for $ZZ$ production at
(n)NNLO accuracy matched to parton showers.  For all (n)NNLO+PS
predictions presented in this paper we make use of the \PYTHIA{8}
parton shower~\cite{Sjostrand:2014zea} with the A14
tune~\cite{TheATLAScollaboration:2014rfk} (\texttt{py8tune 21} in the
input card).  To validate our calculation and to show where shower
effects are crucial, we compare (n)NNLO+PS predictions obtained with
\minnlo{} and (n)NNLO fixed-order predictions obtained with
\Matrix~\cite{Grazzini:2017mhc}.  Additionally, we consider the
inclusion of NLO EW effects.  In the \Matrix{} predictions we set
$\muR=\muF=m_{4\ell}$, and we construct the scale-uncertainty bands
with the same canonical seven-point scale variation used for our
\minlo{} and \minnlo{} results.

Moreover, we compare our predictions with the most recent results by
the CMS collaboration~\cite{CMS:2020gtj} within the fiducial volume
defined in \tab{tab:cuts}, denoted as \setupfiducial. Note that the
reconstructed $Z$ bosons $Z_1$ and $Z_2$ are identified by selecting
the opposite-sign same-flavour (OSSF) lepton pair with an invariant
mass closest to the $Z$-boson mass as $Z_1$ and identifying the
remaining OSSF lepton pair with $Z_2$. Since here we only consider the
different-flavour channel ($e^+e^-\mu^+\mu^-$), the two $Z$ bosons are
unambiguously reconstructed and this procedure only selects which
lepton pair is called $Z_1$ and which $Z_2$. Note that in the
different-flavour channel the additional $m_{\ell^+\ell^-}>4$\,GeV cut
in \tab{tab:cuts} has no effect.  Besides the fiducial setup, we also
consider an inclusive setup (dubbed \setupinclusive), where we only
require a $Z$-mass window between 60\,GeV and 120\,GeV for the two
resonances.

In order to provide the most realistic comparison to experimental
data, our final predictions include effects from hadronization and
multi-particle interactions (MPI).
We also include QED showering effects as provided by \PYTHIA{8}.
In order to prevent charged resonances to radiate photons and photons
to branch into lepton- or quark-pairs, we set the two flags
\texttt{TimeShower:QEDshowerByOther} and
\texttt{TimeShower:QEDshowerByGamma} to \texttt{off}.

Finally, we define dressed leptons by adding to the four-momentum of a
lepton the four-momenta of all photons within a distance $\Delta
R_{\gamma \ell} = \sqrt{\Delta \phi_{\gamma \ell}^2 + \Delta
  \eta_{\gamma \ell}^2} < 0.1$.

\subsection{Integrated cross sections}
\label{sec:minnlo-int}

We start the discussion of our results by first considering integrated
cross sections.  In \mbox{\tab{tab:xsec}} we report predictions both
in the inclusive and in the fiducial setup introduced above for
various perturbative calculations.  Specifically, we consider \minlo{}
predictions, and a number of predictions including NNLO corrections,
both at fixed order and matched to parton showers through \minnlo{}:
besides the complete NNLO predictions (that include the LO
loop-induced $gg$ contribution), we provide the NNLO corrections to
the $q\bar{q}$ channel (dubbed NNLO$_{q\bar{q}}$) and nNNLO cross
sections (as defined before).  For completeness, we also quote nNNLO 
predictions combined with NLO EW corrections, computed at fixed-order with \Matrix{}, either using an additive or multiplicative scheme.  In the
latter predictions we also take into account the photon-induced
contribution at LO and beyond.\footnote{We used the
  \texttt{NNPDF31\char`_nnlo\char`_as\char`_0118\char`_luxqed}~\cite{Manohar:2016nzj,Manohar:2017eqh,Bertone:2017bme}
  PDF set to compute all fixed-order predictions and verified that the
  $q \bar q$ prediction is modified at the permille level with respect
  to the predictions obtained with
  \texttt{NNPDF31\char`_nnlo\char`_as\char`_0118}.}
In order to compare our predictions to fixed-order results,
all \minnlo{} (and \minlo{}) results of \mbox{\tab{tab:xsec}} are obtained at parton
level, without including hadronization, MPI or photon radiation
effects. We have checked explicitly that those effects have a
negligible impact on the integrated cross sections.

The \minnlo prediction and the NNLO result are in excellent agreement
with each other both in the inclusive and in the fiducial setup.
The perturbative uncertainty at (n)NNLO(+PS) is at the 2-3\% level.
In particular, despite the fact that the loop-induced $gg$
process at LO (NLO) contributes only $\sim 6$-$8$\% ($\sim 10$-$15$\%) to the
NNLO (nNNLO) cross section, the uncertainties of the (n)NNLO results are
dominated by the gluon-initated contribution.
The NLO correction for the loop-induced $gg$ channel is particularly
sizable, almost doubling the LO contribution entering at $\as^2$, as
discussed in \sct{sec:gg}.  Accordingly, the nNNLO central prediction
is not included in the NNLO uncertainty band.

The \minlo result is 8-10\% smaller than the \minnlo{} result. Its
uncertainty band, which is considerably larger than the \minnlo one,
does not contain the central (n)NNLO+PS prediction, because scale
variations cannot account for the additional loop-induced $gg$ process
entering at NNLO.  We also note that the \minlo uncertainty band is
larger than the NLO one, and it includes the NLO result. On the
contrary, the NLO uncertainty band is very small and neither \minlo
nor the NNLO central results lie inside it.

\renewcommand{\baselinestretch}{1.5}
\begin{table}[t]
\centering
  \begin{tabular}{l|cc}
    $ \sigma (p p \rightarrow  e^+ e^- \mu^+ \mu^-) $ [fb] & \setupinclusive & \setupfiducial \\
    \toprule
    NLO (\Matrix{}) & 32.50(1)$^{+1.9\%}_{-1.6\%}$ & 16.41(2)$^{+1.9\%}_{-1.6\%}$ \\
    \midrule
    \minlo & 31.42(3)$^{+6.3\%}_{-5.0\%}$ & 16.38(2)$^{+6.0\%}_{-5.0\%}$ \\    
    \midrule
    NNLO$_{q\bar{q}}$ (\Matrix{}) & 34.41(5)$^{+1.0\%}_{-1.0\%}$  &  17.48(2)$^{+1.0\%}_{-1.0\%}$ \\
    NNLO (\Matrix{}) & 36.71(26)$^{+2.4\%}_{-2.1\%}$  &  18.79(11)$^{+2.5\%}_{-2.6\%}$ \\
    nNNLO (\Matrix{}) & 38.24(6)$^{+2.2\%}_{-2.0\%}$ & 19.89(11)$^{+2.6\%}_{-2.8\%}$  \\
    nNNLO+NLO$_{\rm EW}$ (\Matrix{}) & 36.71(26)$^{+2.6\%}_{-2.4\%}$ & 19.06(12)$^{+2.7\%}_{-2.9\%}$  \\
    nNNLO$\times$NLO$_{\rm EW}$ (\Matrix{}) & 35.89(24)$^{+2.5\%}_{-2.3\%}$ & 18.65(11)$^{+2.6\%}_{-2.8\%}$  \\       \midrule
    NNLO$_{q\bar{q}}$+PS (\minnlo)  & 34.36(3)$^{+0.8\%}_{-1.0\%}$ & 17.45(3)$^{+0.9\%}_{-1.0\%}$ \\
    NNLO+PS (\minnlo)  & 36.50(3)$^{+1.9\%}_{-2.0\%}$ & 18.90(3)$^{+2.5\%}_{-2.0\%}$ \\
    nNNLO+PS (\minnlo) & 38.35(3)$^{+2.1\%}_{-2.0\%}$ & 20.04(3)$^{+2.5\%}_{-2.0\%}$ \\
    \midrule
    CMS 13 TeV & $\begin{array}{c} 39.4\pm0.7\text{\scalefont{0.55}(stat)}  \\[-0.15cm] \pm1.1\text{\scalefont{0.55}(syst)} \pm0.9\text{\scalefont{0.55}(theo)} \pm0.7\text{\scalefont{0.55}(lumi)}\end{array}$ & $\begin{array}{c} 20.3\pm0.4\text{\scalefont{0.55}(stat)}  \\[-0.15cm] \pm0.6\text{\scalefont{0.55}(syst)} \pm0.4\text{\scalefont{0.55}(lumi)}\end{array}$\\
  \end{tabular}
  \renewcommand{\baselinestretch}{1.0}
  \caption{Integrated cross sections at various perturbative orders in
    both the \setupinclusive and \setupfiducial region. In brackets we
    report the statistical uncertainties, while scale uncertainties
    are reported in percentages. We also report the inclusive and
    fiducial cross sections measured by the CMS experiment in
    \citere{CMS:2020gtj}.  Since the measured inclusive cross section
    corresponds to on-shell $pp\to ZZ$ production, we have multiplied
    the measured cross section by a branching fraction of
    $\textrm{BR}(Z \rightarrow \ell^+\ell^-) = 0.03366$, as quoted in
    \citere{CMS:2020gtj}, for each $Z$ boson and by a factor of two to
    compare with our predictions for $pp\to e^+e^-\mu^+\mu^-$
    production.  For the measured fiducial cross section the CMS
    analysis includes both different-flavour ($e^+e^-\mu^+\mu^-$) and
    same-flavour ($e^+e^-e^+e^-$, $\mu^+\mu^-\mu^+\mu^-$) decay
    channels of the two $Z$ bosons.  We have therefore divided the
    measured fiducial cross section by a factor of two to compare with
    our $pp\to e^+e^-\mu^+\mu^-$ predictions.}
  \label{tab:xsec}
\end{table}
 \renewcommand{\baselinestretch}{1.0}

Notwithstanding the excellent agreement between the nNNLO(+PS) result
and the fiducial cross section measured by CMS, the theoretical
predictions should be supplemented with EW corrections.
Their inclusion, using either an additive or multiplicative scheme
\cite{Kallweit:2019zez}, has a non-negligible impact on the nNNLO
result and reduces the cross section by about 4-6\% in the
fiducial region, slightly deteriorating the agreement with the
experimental measurement.
We note that EW effects include photon-initiated processes. These have
a negligible impact in the fiducial setup, where the leading lepton
has a transverse momentum larger than 20 GeV, and all leptons have a
transverse momentum larger than 5 GeV. On the contrary, in the
inclusive setup, without a minimal transverse momentum, the
photon-initiated contribution features a collinear divergence.
To avoid this divergence, the CMS analysis~\cite{CMS:2020gtj} imposed
a transverse momentum cut of 5 GeV on the leptons in the evaluation of
the photon-induced component. With this cut, they showed that the
photon-induced contribution is less than 1\% of the total cross
section. For this reason, we set the photon-induced component to zero
for the nNNLO+NLO$_{\rm EW}$ and nNNLO$\times$NLO$_{\rm EW}$ results
in the inclusive case.

\subsection{Differential distributions}
\label{sec:minnlo-diff}

In this section we present our results for differential distributions.
We start by comparing the nNNLO+PS predictions obtained with \minnlo{}
against \minlo and fixed-order nNNLO predictions in the
\setupinclusive{} in \sct{sec:nnlo-minnlo}.  In \sct{sec:minnlodata}
we move to consider the \setupfiducial{} and we compare our \minnlo{}
predictions at nNNLO+PS with the data collected and analyzed by the
CMS experiment~\cite{CMS:2020gtj}.

\subsubsection{Comparison against theoretical predictions}
\label{sec:nnlo-minnlo}

\begin{figure*}[t]
\includegraphics[width=0.49\textwidth]{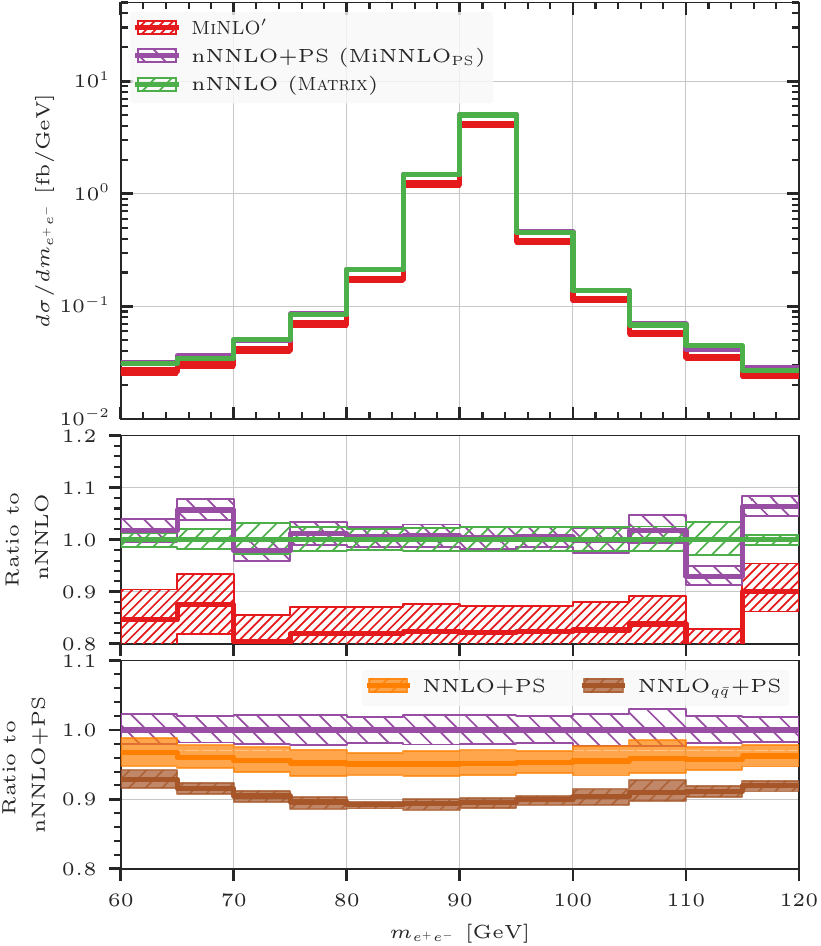}%
\hspace{0.2cm}
\includegraphics[width=0.49\textwidth]{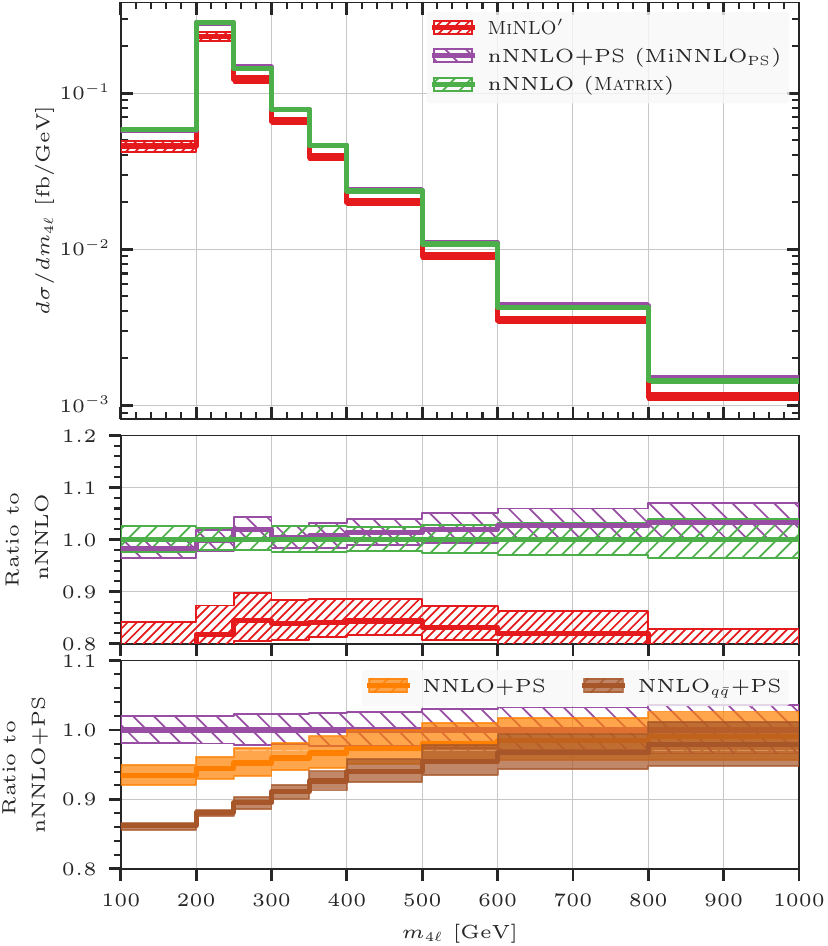}%
\\
\includegraphics[width=0.49\textwidth]{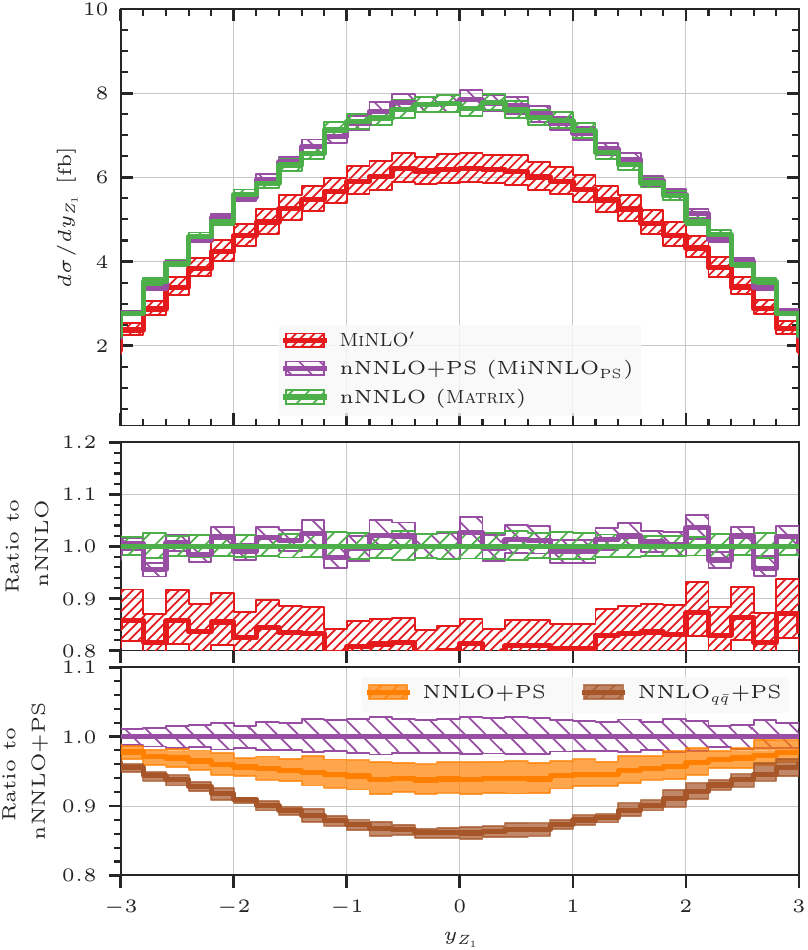}%
\hspace{0.2cm}
\includegraphics[width=0.49\textwidth]{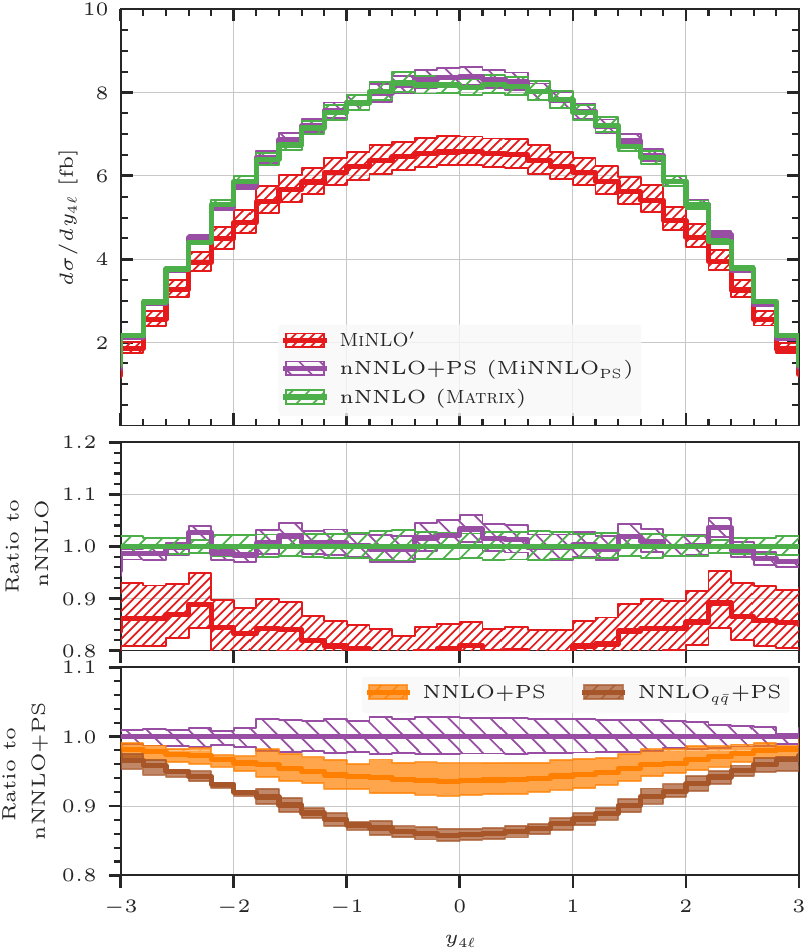}%
  \caption{Comparison between selected distributions computed with
    \Matrix, \minnlo and \minlo. Upper panel: invariant mass of the
    $e^+ e^-$ pair (left) and of the $ZZ$ pair (right); lower panel:
    rapidity of $Z_1$ (left) and of the $ZZ$ pair (right).}
	\label{fig:nnlo-minnlo-comp1}
\end{figure*}

In \fig{fig:nnlo-minnlo-comp1}
we compare nNNLO+PS predictions for \minnlo{} with \minlo{} and nNNLO
predictions at fixed order for four different distributions which are
non-zero at LO.  In particular, we consider the invariant mass of the
$e^+e^-$ pair ($m_{e^+e^-}$), the invariant mass ($m_{4\ell}$) and the
rapidity of the diboson system ($y_{4\ell}$), and the rapidity
($y_{Z_1}$) of the $Z$ boson whose invariant mass is closer to $m_Z$.
We remind the reader that both the \minnlo{} and the \minlo{}
predictions include hadronization and MPI effects, which however are
expected to be very moderate for the inclusive observables considered
in \fig{fig:nnlo-minnlo-comp1}.  We observe a very good agreement
between the nNNLO+PS and the nNNLO predictions, both for the central
values and for the scale-variation bands.  The latter are at the
few-percent level across the whole range shown in the plots, becoming
larger (about $\pm 5$\%) at high $m_{4\ell}$.
Minor differences are visible in the tails of the distributions, in
particular at large $m_{4\ell}$, where the nNNLO-accurate \minnlo{}
and fixed-order predictions however still overlap. Indeed, in the
large invariant-mass region scale choices and terms beyond accuracy
become increasingly important, as it was recently pointed out for
\ww{} production in \citere{Lombardi:2021rvg} and extensively
discussed for $t\bar{t}$ production
\cite{Czakon:2016dgf,Caola:2021mhb}.  The \minlo result is in all
cases about $15$-$20\%$ smaller than the nNNLO results, which provides
mostly flat corrections to the distributions under consideration,
increasing slightly only at large $m_{4\ell}$. We stress that the
relatively flat QCD corrections are a feature of the chosen
distributions (in the inclusive setup) that does not apply in general,
as we shall see below.  Although the \minlo{} uncertainty is a factor
of 5 larger than the \minnlo{} and nNNLO ones, the \minlo{}
predictions do not overlap with the nNNLO-accurate
results. This is not unexpected since a large part of the difference
is caused by the loop-induced $gg$ contribution. Since the latter is
missing in the \minlo{} predictions, the \minlo{} scale variation can
not account for this new production process, which instead
enters the nNNLO results.
From the second ratio panel we can appreciate the effect of the 
loop-induced $gg$ contribution both at LO (comparing NNLO+PS to NNLO$_{q\bar{q}}$+PS) and at NLO 
(comparing nNNLO+PS to NNLO$_{q\bar{q}}$+PS).
It is clear from the plots that due to the gluon flux the impact of the 
loop-induced $gg$ process is more prominent in certain phase-space regions.
The LO (NLO) corrections, which inclusively amount to 
$\sim 6$-$8$\% ($\sim 10$-$15$\%) as pointed out before, 
contribute more significantly in the bulk region of the distributions, 
i.e.\ at the $Z$ resonance in $m_{e^+e^-}$ as well as for small $m_{4\ell}$
and central rapidities.

\begin{figure*}[t!]
\includegraphics[width=0.49\textwidth]{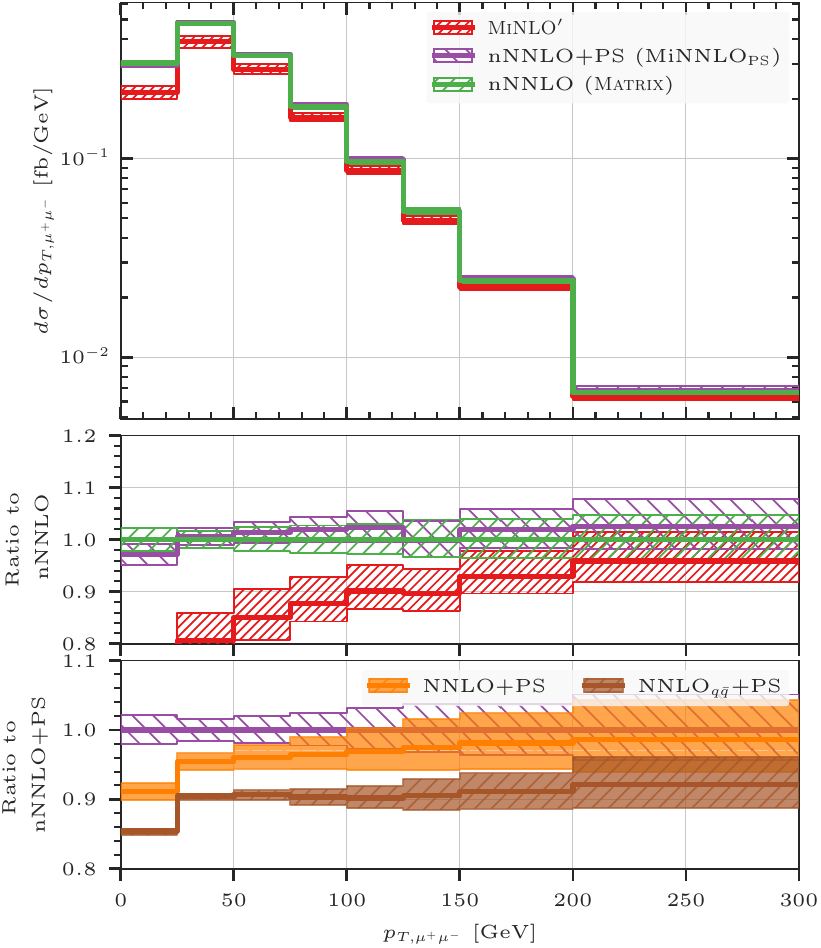}%
\hspace{0.2cm}
\includegraphics[width=0.49\textwidth]{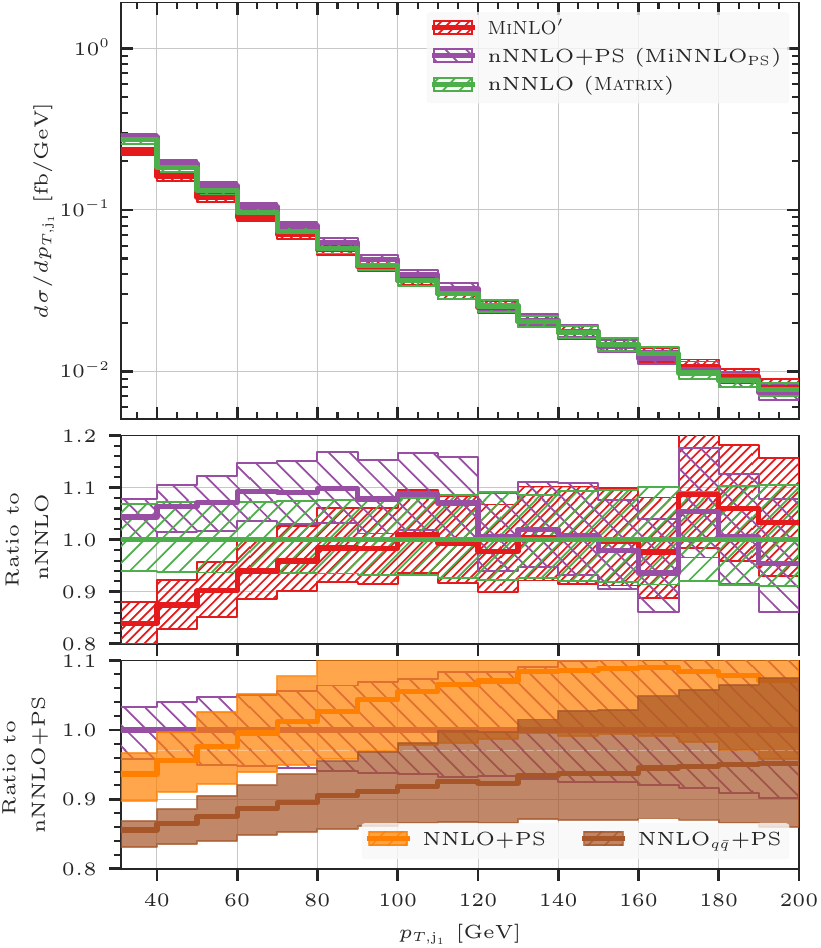}%
  \caption{Same as \fig{fig:nnlo-minnlo-comp1}, for the transverse momentum of the $\mu^+ \mu^-$ pair (left) and of the leading jet (right).}
	\label{fig:nnlo-minnlo-comp2}
\end{figure*}

In \fig{fig:nnlo-minnlo-comp2}
we show the same comparison for the transverse momentum of the $\mu^+
\mu^-$ pair ($p_{T,\mu^+\mu^-}$) and the transverse momentum of the
leading jet ($p_{T,j_1}$) above 30\,GeV.  The first observable is
already defined at LO, while the second one receives its first
contribution only at NLO and its accuracy is thus effectively reduced
by one perturbative order.  The \minnlo and the nNNLO results for
$p_{T,\mu^+ \mu^-}$ are in good agreement with each other in the whole
range shown here. The \minlo result is more than $20\%$ smaller at low
values of the transverse momentum, while it agrees with the other two
predictions at large values of $p_{T,\mu^+ \mu^-}$. Hence, this
distribution shows that in general QCD corrections are not uniformly
distributed in phase space.  By and large, the three predictions for the transverse
momentum of the leading jet display a good agreement,
especially in the tail of the distribution.  The level of agreement
between nNNLO and \minnlo is expected as both predictions are effectively nNLO
accurate at large $p_{T,j_1}$.  The residual scale uncertainties are at the 5-10\% level
and they are larger than those in the other distributions, which is a
direct consequence of the lower accuracy of the predictions for this
distribution.
Looking at the effect of loop-induced $gg$ contribution in the second ratio panel,
we observe a rather peculiar behaviour with the nNNLO+PS corrections being
negative with respect to NNLO+PS for $p_{T,j_1}\gtrsim 80$\,GeV. 
However, this is completely in line
with the results presented in \fig{fig:gg} and it is a consequence of the 
fact that the NNLO+PS predictions include only a LO+PS calculation for the 
loop-induced $gg$ process, which is
not expected to describe the high $p_{T,j_1}$ range 
as it is filled entirely by the parton shower, which has no accuracy in 
this region. This further underlines the need for including NLO corrections 
to the loop-induced $gg$ process. 
Indeed, after including the NLO corrections, the loop-induced
$gg$ contribution reduces
to $5$\% (and less) 
at high $p_{T,j_1}$ (comparing nNNLO to NNLO$_{q\bar{q}}$+PS),
which is more reasonable.

\begin{figure*}[t!]
\includegraphics[width=0.49\textwidth]{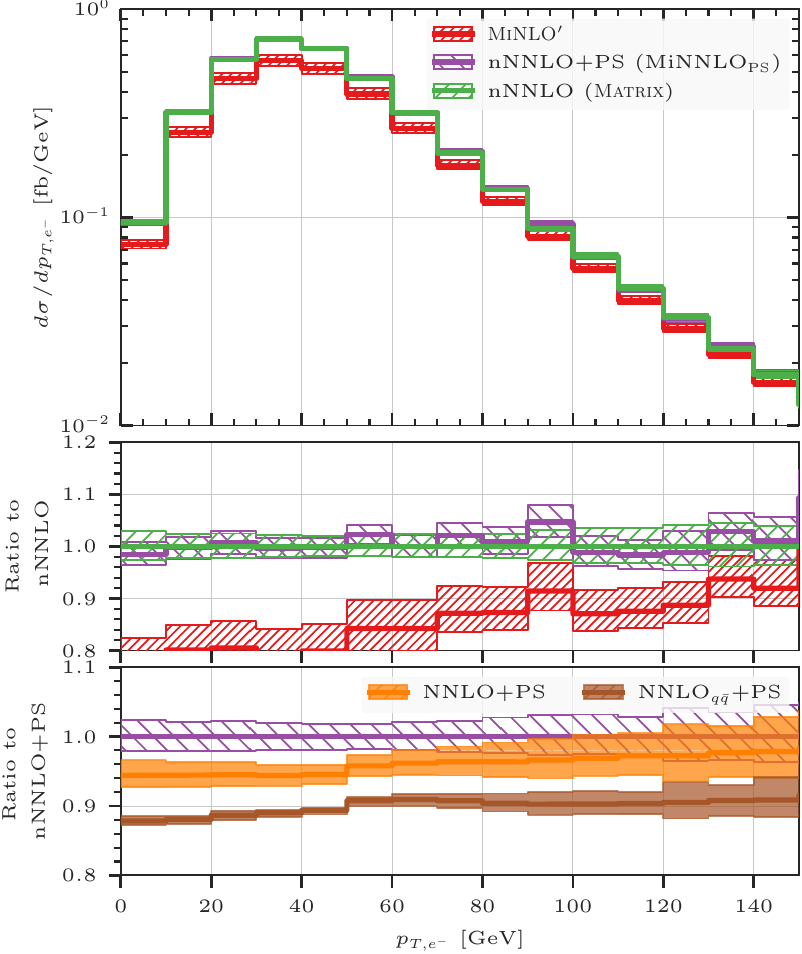}%
\hspace{0.2cm}
\includegraphics[width=0.49\textwidth]{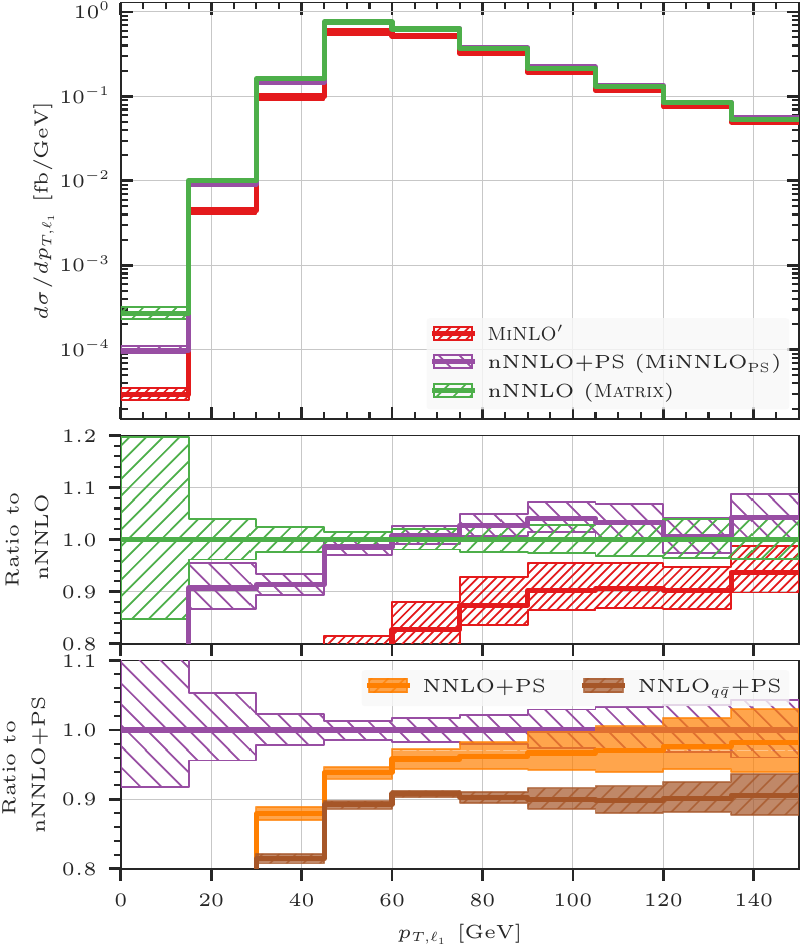}%
  \caption{Same as \fig{fig:nnlo-minnlo-comp1}, for the transverse momentum of the electron (left) and of the leading lepton (right).}
	\label{fig:nnlo-minnlo-comp3}
\end{figure*}

In figure~\ref{fig:nnlo-minnlo-comp3} we show an analogous comparison
for the transverse-momentum spectrum of the electron ($\pte{}$) and of
the leading lepton ($\ptlone{}$).
For the $\pte{}$ distribution we observe excellent agreement over the
whole range between the \minnlo{} and the nNNLO results,
which is fully expected since this distribution should be 
affected very mildly by resummation/shower effects.
We have explicitly checked that a similar level of agreement 
is obtained when considering the same comparison at NNLO$_{q\bar{q}}$ accuracy,
as opposed to the {\sc Geneva} calculation in \citere{Alioli:2021egp},
where differences between the {\sc Geneva} and fixed-order results
are observed for $\pte{}>40$\,GeV. 
When comparing the \minnlo and the \minlo predictions for the $\pte{}$ 
spectrum we observe that
the effect of both the NNLO$_{q \bar q}$ corrections and 
the loop-induced $gg$ contribution is particularly pronounced in the bulk region of
the distribution, where the \minlo result is more than 20\% smaller than
the nNNLO result.
On the other hand, the transverse momentum of the leading lepton is subject to shower
effects, especially at low $\ptlone$, and indeed we observe a difference
between the \Matrix results and the \minnlo predictions
below $40$\,GeV, which become increasingly larger the more steeply
the distribution falls when $\ptlone$ approaches zero.
Above this value, the shower effects are less pronounced and the two
predictions are in good agreement.
By comparing the nNNLO+PS predictions to the NNLO+PS and NNLO$_{q\bar{q}}$+PS results we can see that the impact of the loop-induced $gg$ contribution
is particularly relevant below 40 GeV, and it is also predominantly 
responsible
for the relatively large shower effects that we observe. In fact, we have 
checked that for the NNLO$_{q\bar{q}}$+PS result the 
relative impact of the shower is smaller than for the NLO+PS result in the 
$gg$ channel,
which is expected considering the higher perturbative accuracy (and thereby logarithmic terms) already 
included at fixed order in the $q\bar{q}$ channel.

\begin{figure*}[t!]
\includegraphics[width=0.49\textwidth]{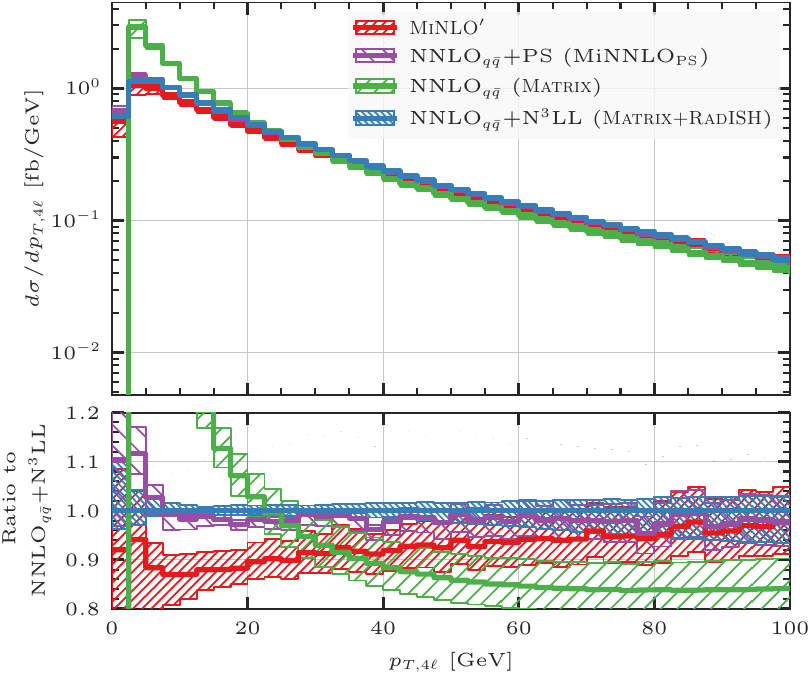}
\hspace{0.15cm}
\includegraphics[width=0.49\textwidth]{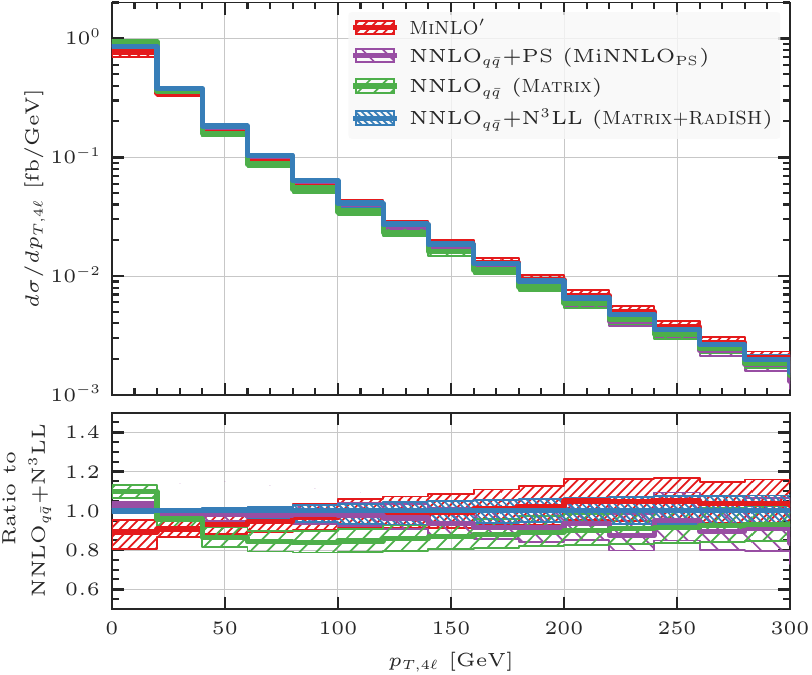}%
  \caption{Same as \fig{fig:nnlo-minnlo-comp1}, for the transverse
    momentum of the $ZZ$ pair for two different ranges of
    $p_{T,4\ell}$. In both plots, we also show the NNLO+N$^3$LL result
    computed with \Matrix{}+\radish~\cite{Kallweit:2020gva}.}
	\label{fig:nnlo-minnlo-comp4}
\end{figure*}

Finally, in \fig{fig:nnlo-minnlo-comp4}
we show predictions for the transverse momentum of the diboson pair
($p_{T,4\ell}$).
In this case, we also show the NNLO+N$^3$LL result obtained with
\Matrix{}+\radish~\cite{Kallweit:2020gva}, which interfaces
\Matrix{}~\cite{Grazzini:2017mhc} to the \radish resummation formalism
\cite{Monni:2016ktx,Bizon:2017rah}, using $\muR=\muF=m_{4\ell}$ and $\Qres=m_{4\ell}/2$ for the resummation scale.
Since \Matrix{}+\radish{} does not include the contribution stemming from the
loop-induced $gg$ channel, we perform this comparison by considering
only the $q \bar q$-initiated process, i.e.\ at the
NNLO$_{q\bar{q}}$(+PS) level.  At small values of the $ZZ$ transverse
momentum we observe an excellent agreement between the NNLO+N$^3$LL
and the \minnlo{} result, especially considering the lower accuracy of
the parton shower in that region; \minnlo{} is between $5$\% and $8$\%
larger than the NNLO+N$^3$LL prediction below 10 GeV and has a larger
uncertainty band reflecting its lower accuracy.  On the other hand,
the \minlo result is $\mathcal O(10 \%)$ smaller than the NNLO+N$^3$LL
and the \minnlo predictions and its uncertainty band does not overlap with
either of the more accurate results below 40\,GeV.  Fixed-order
calculations actually lead to unphysical results in the
small-$p_{T,4\ell}$ region due to large logarithmic corrections, which
need to be resummed to all orders.  Indeed, the NNLO result diverges
at low transverse momentum, and its prediction differs significantly
from the ones including resummation effects.  At larger values
of $p_{T,4\ell}$ the NNLO result is instead in agreement with the
NNLO+N$^3$LL, \minlo and \minnlo predictions, as one may expect since
all of them have the same formal accuracy in the tail of the
distribution.

In conclusion, we observe overall a very good agreement between \minnlo{},
fixed-order, and analytically resummed results across a variety of
distributions, which provides a robust validation of our calculation.
The \minlo{} result, despite the considerably larger uncertainty
bands, rarely overlaps with the (n)NNLO(+PS) predictions, thus
highlighting the importance of higher-order corrections to this
process.  Moreover, certain observables require the resummation of
large logarithmic contributions, which renders the matching to the
parton shower mandatory.

\subsubsection{Comparison against data}
\label{sec:minnlodata}

In this section we compare our \minnlo{} predictions at nNNLO+PS to
the CMS measurement presented in \citere{CMS:2020gtj} in the
\setupfiducial defined in \tab{tab:cuts}.  We have generated the events and estimated the theoretical uncertainties as described in
\sct{sec:setup}.  We remind the reader that our predictions include
MPI and hadronization effects, as well as QED corrections in the shower
approximation.

\begin{figure*}[tbh]
\includegraphics[width=0.48\textwidth]{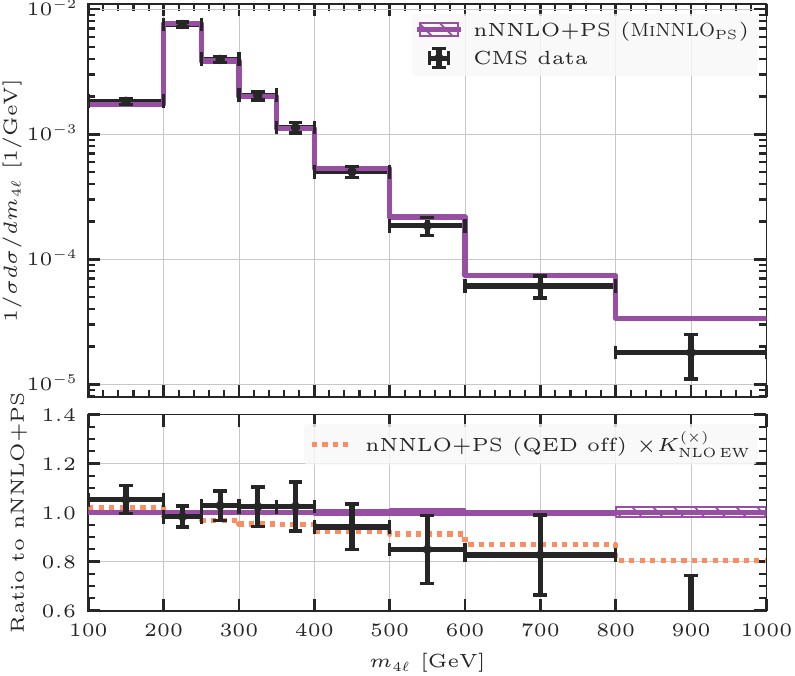}%
\hspace{0.3cm}
\includegraphics[width=0.48\textwidth]{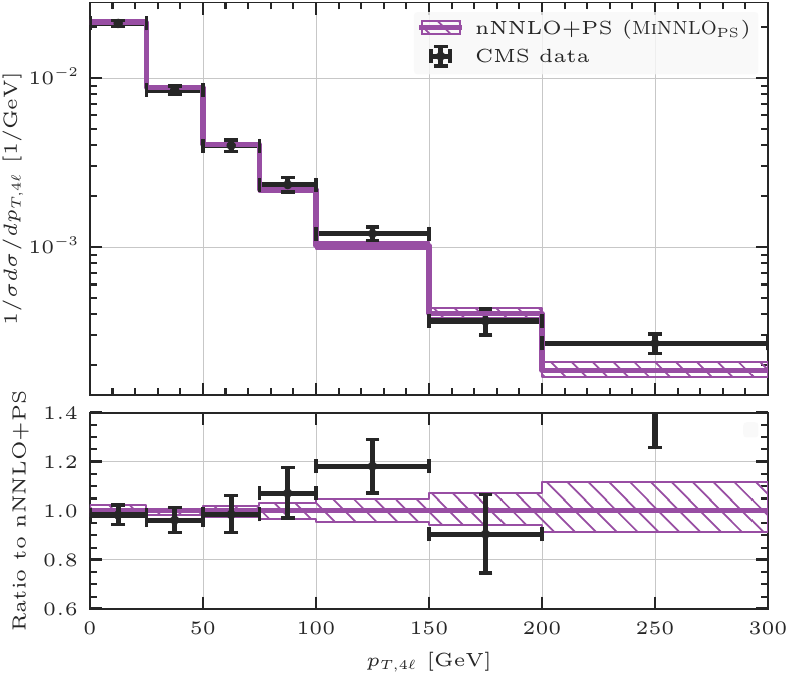}%
\vspace{0.2cm}
\\
\includegraphics[width=0.48\textwidth]{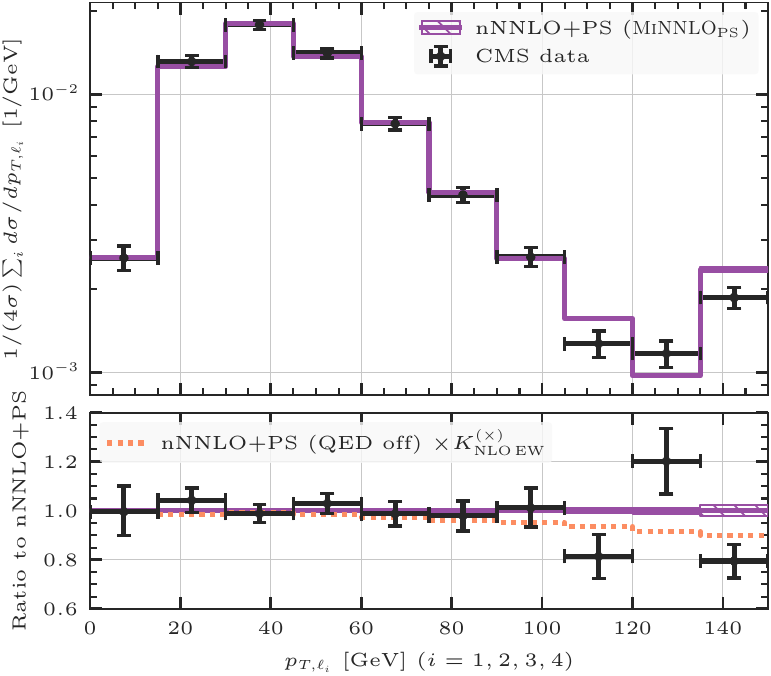}%
\hspace{0.3cm}
\includegraphics[width=0.48\textwidth]{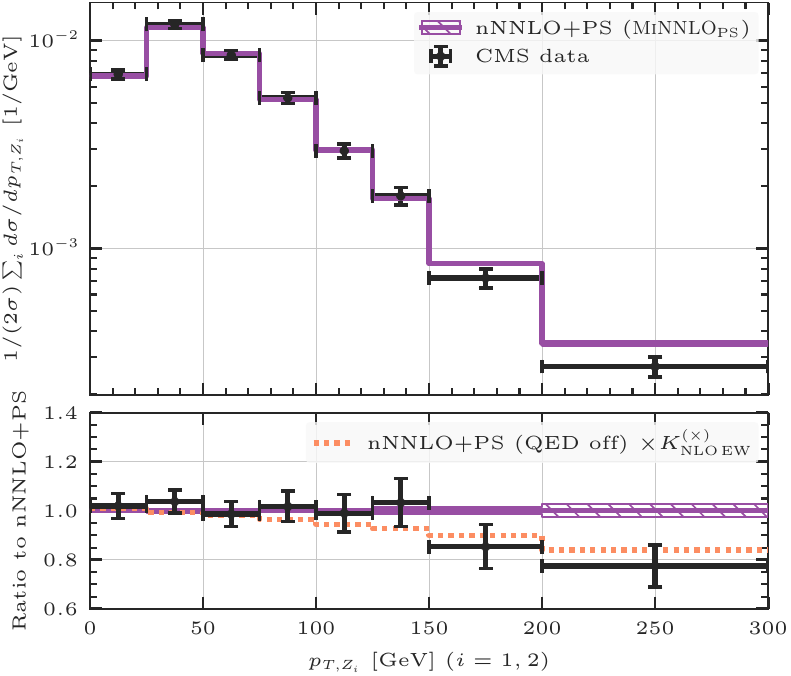}%
\vspace{0.2cm}
\\
\includegraphics[width=0.48\textwidth]{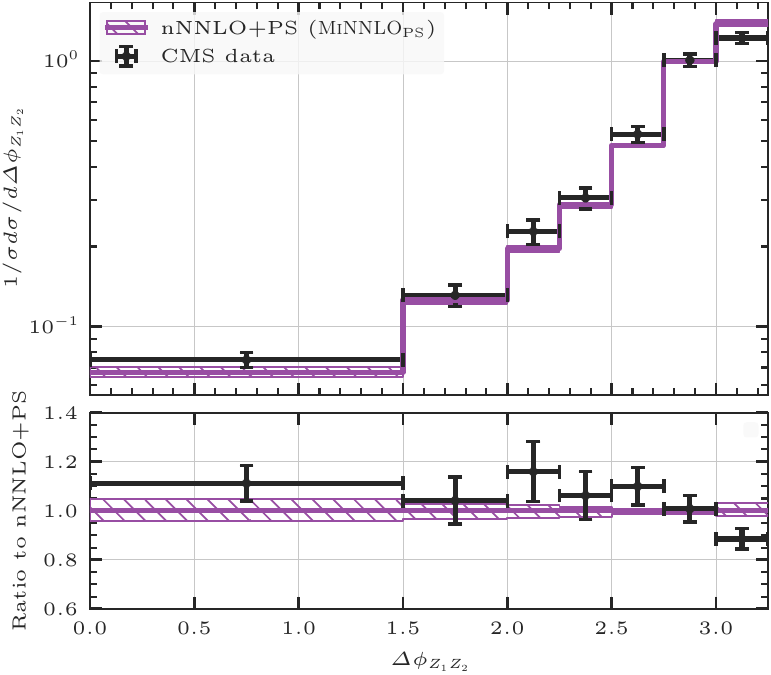}%
\hspace{0.3cm}
\includegraphics[width=0.48\textwidth]{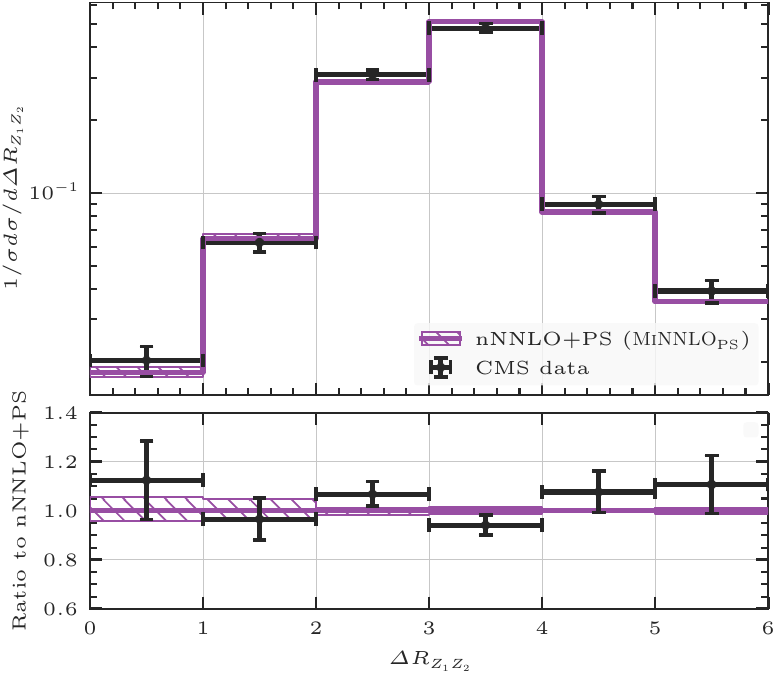}%
  \caption{Comparison between the \minnlo predictions and the CMS data
    of \citere{CMS:2020gtj} based on a 137 fb$^{-1}$ 13\,TeV analysis
    for various observables. The \minnlo predictions include
    hadronization and MPI effects, as well as QED effects as provided
    by the \PYTHIA{8} parton shower. See text for more details.}
	\label{fig:data-minnlo-comp1}
\end{figure*}

The comparison between \minnlo{} predictions and experimental data is
presented in \fig{fig:data-minnlo-comp1}.
Altogether, we show predictions for six observables: the invariant
mass and the transverse momentum of the diboson pair ($m_{4\ell}$ and
$p_{T,4\ell}$), the sum of the four individual transverse-momentum
distributions of each final-state lepton (which corresponds to the
average of the lepton transverse-momentum distributions), the sum of
the two distributions of the transverse momentum of the reconstructed
$Z$ bosons (which analogously corresponds to the average of the $Z$
transverse-momentum distributions), and the separation between the two
$Z$ bosons in the azimuthal angle ($\Delta\phi_{Z_1,Z_2}$) and in the
$\eta$--$\phi$ plane ($\Delta R_{Z_1,Z_2}$).  In all cases, except for
$\Delta\phi_{Z_1,Z_2}$ that has a kinematical endpoint at $\Delta \phi
= \pi$, the last bin shown in the figures also includes the
contribution of the overflow.

By and large, we observe a quite remarkable agreement between our
predictions and the experimental data.  The invariant mass is well
described at low $m_{4\ell}$, but there is a tendency of the data to
undershoot the prediction at large $m_{4\ell}$, 
with the last bin being almost two standard
deviations away. In this region EW corrections are known to be important
and they are only partly included here through the QED shower. Below, we 
discuss how the inclusion of the NLO EW corrections at
fixed order improves the agreement with data in this
region. 
The transverse-momentum distribution of
the $ZZ$ pair is also well described, except for a two-sigma deviation
in the last bin, with a remarkable agreement for $p_{T,4\ell}$ values
below $ \sim 100$\,GeV, where the all-order corrections provided by
the shower are particularly important.  The two averaged distributions
of $p_{T,\ell_i}$ and $p_{T,Z_i}$ also compare very well to \minnlo{},
with  deviations  in the tail of the distributions only.
In the last bins the
experimental data are about two standard deviations away from
the theoretical predictions, which can again be  related to the missing 
EW corrections, as discussed below. 
The $\Delta \phi_{Z_1,Z_2}$ and the
$\Delta R_{Z_1,Z_2}$ distributions are also  very well described by
\minnlo{}, with the data fluctuating (within one sigma, except for one
bin with a two-sigma deviation) around the central theoretical
prediction across the whole plotted range.

The comparison at the level of integrated cross section in
\sct{sec:minnlo-int} showed that the inclusion of NLO EW effects has a
small, but non-negligible impact in the fiducial setup.  Since in our
comparison with data we include QED effects via parton-shower
matching, one may wonder whether the proper inclusion of NLO EW
effects in a Monte Carlo context, see
e.g.~\citere{Brauer:2020kfv,Chiesa:2020ttl}, would further improve the
agreement with the data, especially in the tails of distributions
where EW logarithms are important.  A possible way to assess the
impact of the EW corrections beyond the parton shower approximation is
to apply to the \minnlo predictions a differential $K$-factor correction 
for the NLO EW corrections that is computed at fixed order accuracy.

We have done this exercise turning off the QED shower in
the \minnlo predictions to avoid double counting.  The central
rescaled prediction is shown in the lower ratio panels in
\fig{fig:data-minnlo-comp1}.  We adopt as our default a
factor $K^{(\times)}_{\rm NLO EW}$, defined using the multiplicative scheme
nNNLO$\times$NLO$_{\rm EW}$ \cite{Kallweit:2019zez}, which includes an estimate of mixed higher-order corrections, 
divided by the nNNLO
result. 
Note that for distributions starting at NLO
QCD we do not perform this additional comparison, since one would need
to compute the EW corrections to the $ZZ$+1-jet process.  We find that
the inclusion of NLO EW corrections within this approximation improves the
agreement with the experimental data for the tails of the $m_{4 \ell}$
and the averaged $p_{T,Z_i}$ distributions, where the effects of Sudakov
logarithms are expected to be visible.  For the averaged $p_{T, \ell_i}$ distribution
their impact is a bit milder, also because the distribution extends to lower 
values, and there is no significant improvement compared to data.
We leave a consistent inclusion of NLO EW effects in our \minnlo
predictions with a complete and consistent matching to QCD and QED
showers to future work.

\section{Conclusions}
\label{sec:summary}

In this work, we have advanced the state-of-the-art for Monte Carlo
simulations of $Z$-boson pair production at the LHC. For the $q\bar{q}$-initiated process we have
matched NNLO QCD predictions to parton showers using the \minnlo method.
We have included the loop-induced $gg$-initiated process, which contributes at 
${\cal O}(\alpha_s^2)$, in the \POWHEGBOXRES{} framework
at NLO QCD 
accuracy matched to parton showers. When combined, the ensuing 
nNNLO+PS results
constitute the most accurate theoretical predictions for this process to date.
We remind the reader that the benefits of the \minnlo approach reside
in (1) the possibility to include NNLO corrections on-the-fly, without
the need of any a-posteriori reweighting, (2) the absence of any
merging scale or unphysical boundaries to partition the phase space
into different regions according to the number of resolved emissions,
and (3) the fact that the logarithmic accuracy of the parton shower is
preserved by the matching (when using a $p_T$-ordered shower). 
We stress that this last feature is in general far from trivial.

We have performed an extensive comparison of our \minnlo{} predictions 
against (n)NNLO fixed-order results and the analytic resummation in the 
transverse momentum of the four-lepton system. We found excellent
agreement with fixed-order predictions in phase-space regions 
where shower effects are expected to be small. 
As expected, for distributions that have a singularity at fixed order the \minnlo predictions feature the appropriate Sudakov damping close
to the singularity and yield physical results.
In particular, the comparison 
to the NNLO+N$^3$LL $p_{T,4\ell}$ spectrum showed quite a remarkable 
agreement.
Moreover, we have shown that \minnlo{} corrections are 
at the level of $15$-$20$\% with respect to \minlo{}, and that the matching 
to the parton shower is crucial for observables sensitive to soft-gluon effects.
It is interesting to notice that we do not observe the mild tension in 
the $\pte{}$ distribution observed in \citere{Alioli:2021egp}
when comparing NNLO$_{q\bar{q}}$+PS to fixed-order predictions, which motivates a
more comprehensive comparison between {\sc Geneva} and \minnlo{} 
predictions in the future.

We have compared our nNNLO+PS predictions against 13 TeV CMS data of
\citere{CMS:2020gtj} and found excellent agreement both at the level
of production rates and shapes of kinematical distributions, with
nNNLO+PS predictions and CMS data agreeing on almost all bins within
one sigma. In the few bins where the differences are at the two-sigma level
we have shown that the inclusion of NLO EW corrections removes those differences
in most instances.
Our final results have missing higher-order uncertainties that are of
the order of 2\% both for inclusive and fiducial cross sections. These
uncertainties are of similar size as the current precision of experimental results,
which will further decrease in the future. 
It is then clear that theoretical predictions with an accuracy
comparable to that of the results presented in this paper are
mandatory to fully exploit \zz{} cross section measurements at the
LHC. This is particularly the case when using \zz{} production data for
off-shell Higgs cross section measurements in the $H\to 4\ell$ channel
to put bounds on the Higgs-boson width, or when constraining
the coefficients of effective-field-theory operators or
anomalous triple-gauge couplings.
It is interesting to note that, even though the loop-induced $gg$
contribution is only about 10\% of the total cross section, its
uncertainty dominates our final predictions. It is unlikely that a
  three-loop calculation for this process will become available in the
  near future.
Still, the theoretical precision can be further improved if one
imposes fiducial cuts that suppress the loop-induced gluon fusion
contribution. For instance, when requiring a large invariant mass of
the lepton system or when considering high-$p_T$ leptons, as done 
in BSM searches, the gluon-fusion
contribution becomes less important since the gluon PDFs decrease 
strongly at
large $x$ values.
On the other hand, electroweak effects become more important in these
regions. The approximate combination considered here,
which already increases the agreement 
with the experimental data in the high-energy tails,
needs to be improved by combining highest-order 
QCD and QED corrections consistently
in parton-shower simulations in the future.

Since the evaluation of the two-loop contributions is numerically highly
demanding, we made full use of the reweighting facility of \POWHEG{}
and introduced the possibility to evaluate the two-loop contributions
only at the very end of the event generation, considerably 
speeding up the calculation.
The code used for our simulations will
be soon made publicly available within \POWHEGBOXRES{}.
We are confident that this will be valuable
for upcoming experimental measurements of $ZZ$ production at the LHC, which require
an accurate and fully exclusive simulation of hadron-level events, including
all-order, non-perturbative, and QED effects.

\section*{Acknowledgments}
We are grateful to Stefano Forte, Pier Francesco Monni, Paolo Nason
and Emanuele Re for various fruitful discussions.  We thank the
authors of \citere{Alioli:2021wpn} for providing the {\tt gg4l} code
and especially Jonas Lindert for very helpful communication. 
We have used the Max Planck Computing and Data Facility (MPCDF) in Garching 
and the cloud-computing facilities of the group of Massimiliano Grazzini
 at the University of Zurich
to carry out the simulations related to this work. 
LB and LR are supported by the Swiss National Science Foundation (SNF)
under contract 200020\_188464.

\bibliography{MiNNLO}
\bibliographystyle{JHEP}

\end{document}